\documentclass[trackchanges]{aastex701}

\usepackage{placeins}  
\usepackage{float}
\usepackage{url}

\usepackage{graphicx}
\usepackage{color}
\usepackage{siunitx}
\usepackage{hyperref}
\usepackage{xcolor}
\usepackage{rotating}
\usepackage{longtable}
\usepackage{booktabs}
\hypersetup{
	colorlinks,
	citecolor={blue},
	urlcolor={red},
	linkcolor={blue},
}

\usepackage{float}
\begin{document}

\title{Effect of Solar Flares on Decayless Kink Oscillations in Nearby Coronal Loops}

\author{Zhiyi Li}
\affiliation{Faculty of Information Engineering and Automation, Kunming University of Science and Technology, Kunming, Yunnan 650500, China}
\email{2950438261@qq.com}

\author[orcid=0000-0001-6423-8286]{Valery M. Nakariakov}
\affiliation{Centre for Fusion, Space and Astrophysics, Department of Physics, University of Warwick, CV4 7AL, Coventry, UK}
\affiliation{G-LAMP NEXUS Institute/School of Space Research, Kyung Hee University,  Yongin, 17104, Republic of Korea}
\affiliation{Centro de Investigacion en Astronomía, Universidad Bernardo O’Higgins, Avenida Viel 1497, Santiago, Chile}
\email{V.Nakariakov@warwick.ac.uk}

\author[orcid=0000-0002-9514-6402]{Ding Yuan}
\affiliation{Key Laboratory of Solar Activity and Space Weather, School of Aerospace, Harbin Institute of Technology, Shenzhen, Guangdong 518055, China}
\email[show]{yuanding@hit.edu.cn}

\author[orcid=0000-0003-4709-7818]{Song Feng} 
\affiliation{Faculty of Information Engineering and Automation, Kunming University of Science and Technology, Kunming, Yunnan 650500, China}
\email[show]{feng.song@kust.edu.cn}

\begin{abstract}
We present a statistical study of 130 solar flares (B to X class) that lack soft X-ray quasi-periodic pulsations and show no kink oscillations of nearby coronal loops visible in SDO/AIA 171~\AA~images. The aim is to investigate whether decayless kink oscillations of coronal loops respond to nearby flaring activity. Using the Fractional Anisotropy-based Video Motion Magnification technique, we detected low-amplitude decayless oscillations in all 130 loops before, during, and after each flare, confirming their ubiquitous nature. Oscillation periods are found to range from 122~s to 268~s, and the projected displacement amplitudes are 0.023--0.111~Mm. No amplitude--period correlation is found. For each event, we estimated the amplitude before, during, and after the flare. Across all flare classes, the average amplitude remains unchanged. However, in some specific cases, the oscillation amplitude may exhibit minor changes. For B-, C-, and M-class flares, the fraction of events with an amplitude change exceeding 10\% is approximately 23\%, 41\%, and 36\%, respectively. In M-class flares, such minor amplitude increases occur four times more often than decreases; in X-class flares (only six events), decreases dominate by a factor of three. 
The fraction of events that exhibit an increase in the amplitude of more than 20\% appears to be highest when the loop centre is located at a distance of 100--120~Mm from the flare site, reaching 33\% (6 out of 18 events).
Overall, the amplitude of decayless kink oscillations does not undergo a major change in response to nearby flares, especially for less powerful classes, suggesting that flare-related processes such as blast waves and reconnection inflows have little effect on the energy supply to oscillating loops.
\end{abstract}
\keywords{\uat{Solar coronal waves}{1995}  --- \uat{Solar coronal seismology}{1994} --- \uat{Solar flares}{1496} }

\section{Introduction} 
\label{sec:int}

Kink oscillations of coronal loops remain one of the most studied oscillatory phenomena in the solar corona \citep[e.g.,][]{2021SSRv..217...73N}. Kink oscillations appear as periodic transverse displacements of coronal loops from equilibrium. Typically, all segments of the loop oscillate either in phase or in anti-phase with each other, resembling oscillation of a guitar string. In the vast majority of cases, only the fundamental standing harmonic is detected, with the maximum displacement at the loop top and two nodes, at the footpoints. Typical oscillation periods range from a few tens of seconds to a few tens of minutes \citep[e.g.,][]{2015A&A...583A.136A, 2019ApJS..241...31N, 2023ApJ...944....8L}, which allows for the confident detection with existing EUV imagers \citep[e.g.,][]{1999ApJ...520..880A, 2019ApJS..241...31N, 2023ApJ...944....8L}. The oscillation period scales almost linearly with the length of the loop \citep[e.g.,][]{2016A&A...585A.137G, 2015A&A...583A.136A, 2023ApJ...944....8L, 2024A&A...690L...8L}. 

Kink oscillations are known to appear in two regimes. Decaying kink oscillations are characterised by maximum transverse displacement amplitudes of about 1--10~Mm, which is comparable to or exceeding the minor radii of oscillating loops \citep{2019ApJS..241...31N}. In this regime, the oscillations are usually excited by an initial displacement of the loop from equilibrium by a low coronal eruption \citep{2015A&A...577A...4Z, 2022SoPh..297...18Z, 2025SoPh..300...81L}, e.g., a jet \citep{2023ApJ...955...89Q}. There is also observational evidence for the excitation of decaying kink oscillations by coronal rain \citep{2017A&A...601L...2V}. Several other excitation mechanisms have been proposed. Among them are the initial displacement of the loop by a blast wave initiated by a nearby flare \citep[e.g.,][]{2009SSRv..149..153O, 2004ApJ...614L..85H,2023A&A...675A.169L,2024ScChE..67.1592L},
and the loop contraction due to a post-flare implosion \citep{2013ApJ...777..152S,2017A&A...607A...8P}.

In the decaying regime, kink oscillations show rapid damping, within a few to several oscillation cycles \citep{2019ApJS..241...31N}. The damping time has been found to scale with the oscillation period \citep{2002ApJ...576L.153O, 2013A&A...552A.138V, 2016A&A...585A.137G, 2019ApJS..241...31N}. The damping is attributed to resonant absorption, which corresponds to the linear conversion of a collective kink oscillation into unresolved torsional motions within the loop \citep[e.g.,][]{2002ApJ...577..475R, 2004ApJ...606.1223V, 2020SSRv..216..140V}. However, observational evidence of the dependence of the damping effectiveness on the oscillation amplitude points to the nonlinear nature of the damping process \citep{2016A&A...590L...5G}. A nonlinear mechanism for the damping of kink oscillations, which has been intensively studied theoretically, is the Kelvin–Helmholtz instability \citep[e.g.,][]{2008ApJ...687L.115T, 2014ApJ...787L..22A, 2016A&A...595A..81M, 2024A&A...688A..80K}, although its occurrence has not yet been unequivocally confirmed observationally.

The other regime, that of decayless kink oscillations, is characterised by an almost stationary amplitude \citep{2012ApJ...751L..27W, 2013A&A...560A.107A, 2021SSRv..217...73N}. The displacement amplitude is much lower than in the decaying regime, typically less than 1 Mm. Decayless kink oscillations have been detected to last up to several tens of oscillation cycles, without a systematic evolution of either the oscillation parameters or the loop itself \citep[][]{2022MNRAS.513.1834Z}.
Despite the low, often sub-pixel amplitude, the simultaneous detection of decayless kink oscillations with two different EUV imagers rules out an artificial origin for this phenomenon \citep[][]{2022MNRAS.516.5989Z}. Furthermore, the persistent oscillatory Doppler shift of coronal lines near loop tops confirms the natural origin of decayless kink oscillations \citep{2012ApJ...759..144T}. The detection of the fundamental and second harmonics which appear simultaneously \citep{2018ApJ...854L...5D} further supports this conclusion.

The decaying and decayless kink oscillation regimes appear in the same loop.  
A loop has been found to oscillate consecutively in decayless, decaying, and then again decayless regimes, all with the same oscillation period \citep{2013A&A...552A..57N}. 
This finding suggests that the same damping mechanism operates in both regimes, but that in the decayless regime it is somehow counteracted by an energy supply. Revealing this energy supply mechanism is of great interest in the context of the coronal heating problem \citep[e.g.,][]{2019FrASS...6...38K, 2020SSRv..216..140V, 2023ApJ...952L..15L} and is therefore the subject of extensive study.

The absence of a pronounced peak in the distribution of decayless kink oscillation amplitudes against the oscillation period rules out energy supply via the leakage of 3‑min or 5‑min oscillations from the lower layers of the solar atmosphere into the corona \citep{2016A&A...591L...5N}. There has been detected a case where decayless kink oscillations appear to co-exist with quasi-periodic pulsations (QPP) of the non-thermal emission produced by a flare occurring in the vicinity of the oscillating loop \citep{2023A&A...680L..15L,2020ApJ...893L..17L}. However, in that case, QPP could be induced by the kink oscillation \citep[see, e.g.,][]{2006A&A...452..343N}. 
Hence, in general, the energy supply must be aperiodic in nature. 

Theoretical modelling has demonstrated that a decayless oscillation with the natural kink period can be produced by random motions of the footpoints, e.g., by granulation \citep{2020A&A...633L...8A, 2021MNRAS.501.3017R,  2024A&A...681L...6K}. However, this scenario is poorly consistent with the linear polarisation of decayless kink oscillations, established by their observations from two non-parallel lines of sight \citep{2023NatCo..14.5298Z}. 

The self-oscillatory, or \lq\lq violin\rq\rq\ mechanism for decayless kink oscillations proposes that the oscillations are sustained by a continuous, steady energy flow from an external driver, such as quasi-steady coronal or footpoint motions \citep{2016A&A...591L...5N}. This scenario is analogous to the oscillations of a violin string in response to a quasi-steady motion of the bow. The plausibility of this mechanism has been confirmed by full-MHD numerical simulations \citep{2020ApJ...897L..35K}. The self-oscillatory mechanism may be linked with the random driver mechanism, provided the random process has a red (or similar) noise spectrum \citep{2025ApJ...993L..35Z}. In that case the photospheric motions are dominated by large-scale motions, in comparison with the kink oscillation period, which can therefore be considered quasi-static. Furthermore, the self-oscillatory mechanism is supported by combined analyses of coronal kink oscillations and photospheric motions \citep{2025A&A...696A.125P}. In addition, the damping envelopes of decaying kink oscillations are more consistent with a super-exponential transition to the decayless regime, suggested by the self-oscillatory model, than with exponential damping \citep{2024MNRAS.531.4611N}.

The observational detection of decayless kink oscillations characterised by small (often $<0.1$~Mm) amplitudes intrinsic to this regime, benefits from the application of video motion magnification techniques. \citet{2016SoPh..291.3251A, 2021SoPh..296..135Z} adapted the motion magnification technique that advanced the Eulerian video magnification method originally developed by \citet{Wu2012, Wadhwa2013}, for the detection of kink oscillations with sub-pixel amplitudes in time sequences of EUV images of the corona. Another technique is the Fractional Anisotropy-based Video Motion Magnification (FA-VMM) \citep{li2025revealing}. This method reliably distinguishes genuine physical motion from noise through fractional anisotropy (FA) analysis, significantly suppressing background noise amplification. Simultaneously, it incorporates a Hierarchical Edge-Aware Regularisation (HEAR) module to effectively avoid artefacts such as blurring and ripples, while preserving the sharpness and phase coherence of oscillatory structures. The FA-VMM technique thus exhibits excellent performance in noise suppression, structural fidelity, and artefact control.
Consequently, the FA-VMM technique effectively distinguishes genuine physical motion from noise by analysing the anisotropic characteristics of motion signals \citep{Takeda2019VideoMI}, enabling high-fidelity amplification of oscillatory displacements in the plane of the sky in time sequences of images (typical amplification factors of 5–35). This provides a reliable tool for the systematic detection of sub-pixel decayless kink oscillations \citep{li2025revealing}.

So far, statistical studies of decayless kink oscillations have been restricted to periods of quiet activity periods. There have been only a few previous case studies of decayless kink oscillations during flares \citep{2021A&A...652L...3M, 2022RAA....22j5017S}.
For further constraining our understanding the physical mechanisms responsible for decayless kink oscillations it is of interest to investigate whether the amplitude of the oscillations shows any systematic variation during major energy releases occurring nearby, when coronal conditions may experience a significant change. In this study, we estimate the kink oscillation amplitudes in the vicinity of solar flares. The chosen flares do not have statistically significant QPP which could periodically drive kink oscillations, and do not excite decaying kink oscillations. Decayless kink oscillations in loops situated near the flare epicentre are detected with the use of the FA-VMM technique. 
In this study, we restricted our analysis to the dynamics of a single, isolated, and well-contrasted loop in the vicinity of each flare. This limitation is justified by the need to detect subtle changes in the sub-pixel displacement amplitude, which are sensitive to the background conditions of the candidate loop. In most cases, it is not straightforward to identify more than one loop suitable for this type of analysis in an event. On the other hand, we analysed decayless kink oscillations across a large number of events spanning various flare classes, which distinguishes our work from the case studies by \citet{2021A&A...652L...3M} and \citet{2022RAA....22j5017S}.
The paper is organised as follows. In Section~\ref{sec:d_and_m} we describe the analysed data and applied methodology. The results obtained are presented in Section~\ref{sec:res}, and discussed and summarised in Section~\ref{sec:con}.  

\section{Data and Methods}
\label{sec:d_and_m}
\subsection{Data Selection and Preprocessing}
\label{sec:sel}
The flare sample in this study was selected from an event pool classified as “SXR non-QPP” through a systematic analysis using a Bayesian Markov Chain Monte Carlo approach \citep{2023ApJ...944...16G}. This classification is based solely on the analysis of light curves in the GOES soft X-ray band and does not involve other wavelengths. From this pool, we further identified 160 events in which the associated coronal loops exhibited no decaying periodic transverse displacements, i.e., decaying kink oscillations in the raw SDO/AIA 171~\AA\ images. The flare class distribution of the sample is as follows: 60 B-class, 55 C-class, 37 M-class, and 8 X-class flares. All data sequences have a uniform temporal resolution (12 seconds) and spatial resolution (0.6 arcseconds per pixel). Each event was observed for at least 30 minutes before the flare onset to at least 30 minutes after its end. To ensure precise motion analysis, all images were preprocessed with solar rotation correction, differential rotation compensation, and cross-correlation alignment to eliminate effects resulting from solar rotation and instrumental jitter.

\subsection{FA-VMM and Parameter Extraction}
\label{sec:favmm}
We employ the validated Fractional Anisotropy-based Video Motion Magnification (FA-VMM) technique to amplify the transverse displacement of coronal loops, selecting a magnification factor of 20. This factor has been verified through synthetic data testing to fall within the linear response range of the technique, allowing high-fidelity motion enhancement (\cite{li2025revealing} for detailed validation).

For each event, we selected coronal loops with clear morphology, well-defined boundaries, and continuous traceability in AIA 171~\AA\ images for analysis. Their spatial locations were manually tracked to define the slice orientation and generate time--distance maps. Subsequently, by visually inspecting the magnified sequence, the loop's transverse displacement was manually marked to obtain its oscillatory displacement time series. Crucially, as all samples were selected from events with oscillations invisible to the naked eye in the original images, none of the extracted displacement sequences exhibited discernible decaying features. Quantitative verification using a damped harmonic oscillator model confirmed that the damping time is either unconstrained or significantly exceeds the observation window, justifying the use of the simple harmonic model. Consequently, we approximated the oscillations by a harmonic function 
\begin{equation}
y(t) = A_0 
\sin\left(\frac{2\pi}{P} t + \Phi\right
),
\label{eq:sine_model}
\end{equation}
where $A_0$, $P$ and $\Phi$ correspond to the oscillation amplitude, period, and phase, respectively. Assuming that the oscillation period and amplitude may vary with time, each fit was based on the use of a signal of about 600~s long. 
The fitting process used a nonlinear least-squares method, with errors of the best-fitting parameters estimated from the covariance matrix.

To investigate the transient effects of solar flares and the subsequent evolution of the oscillation amplitude, we selected 130 events from the total sample that exhibited high-quality oscillatory patterns which exhibit at least three cycles of oscillations, across the pre-, during-, and post-flare phases for in-depth analysis. The flare start and end times were taken from the NOAA/GOES X-ray flare catalog. Based on these times, each flare event was divided into three phases for analysis: the pre-flare phase (20 minutes prior to onset), the during-flare phase (from onset to end), and the post-flare phase (20 minutes after the end of the flare).

\section{Results}
\label{sec:res}
After applying the FA-VMM technique to 160 imaging data cubes which included the flare sites and time intervals, we successfully extracted decayless kink oscillation signals from the associated coronal loops in all cases (160/160).

\subsection{Examples of detected decayless kink oscillations}
\label{sec:exa}

\begin{figure*}[ht!]
\centering
\includegraphics[width=\textwidth]{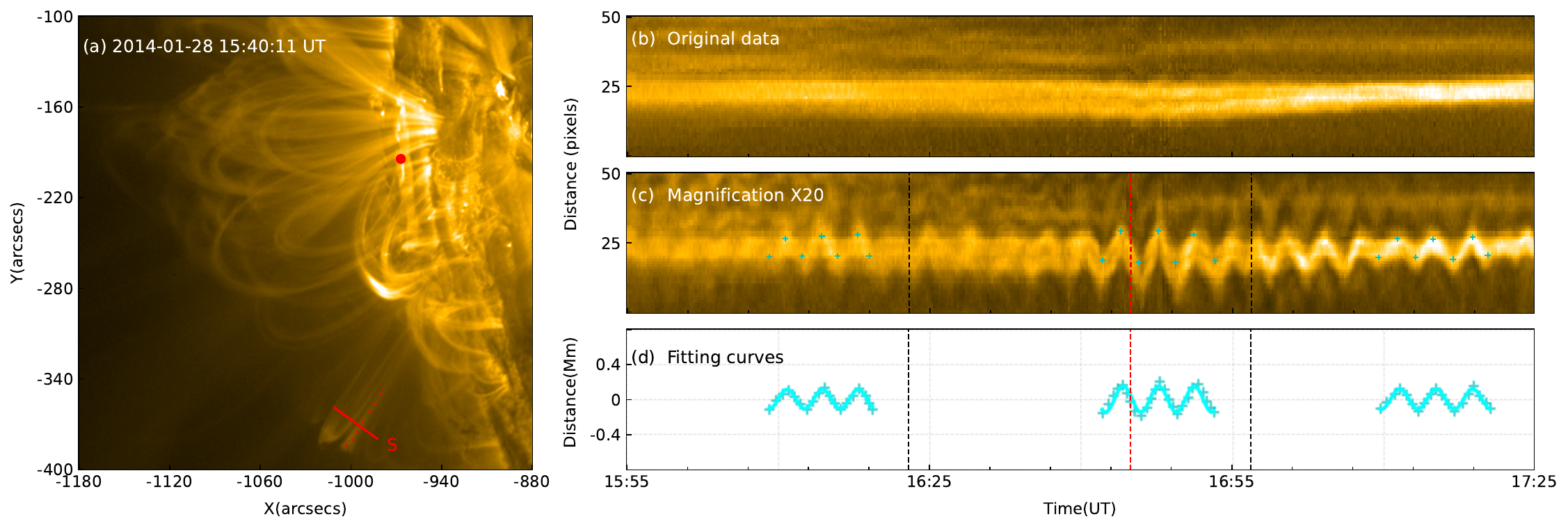} 
\caption{C8.5-class flare event on January 28, 2014. (a) Target observation region [-1180, -880, -400, -100] (arcsec), with the location of a coronal loop of interest marked by the red dashed line, and the slice used in the construction of the time--distance map marked by the red solid line. {The red solid dots mark the positions of the flare centre.} (b) Raw coronal loop time--distance map. (c) Coronal loop time--distance map amplified 20 times using the FA-VMM technique. (d) Fitted oscillation curve. The cyan "+" mark the instantaneous displacements of the oscillating loop, and the cyan solid line represents the fitted decayless harmonic oscillation curve. The black vertical dashed lines mark the start and end times of the flare, and the red vertical dashed line marks the peak time of the flare.}
\label{fig:2014-01-28-C8.5}
\end{figure*}

\begin{figure*}[ht!]
\centering
\includegraphics[width=\textwidth]{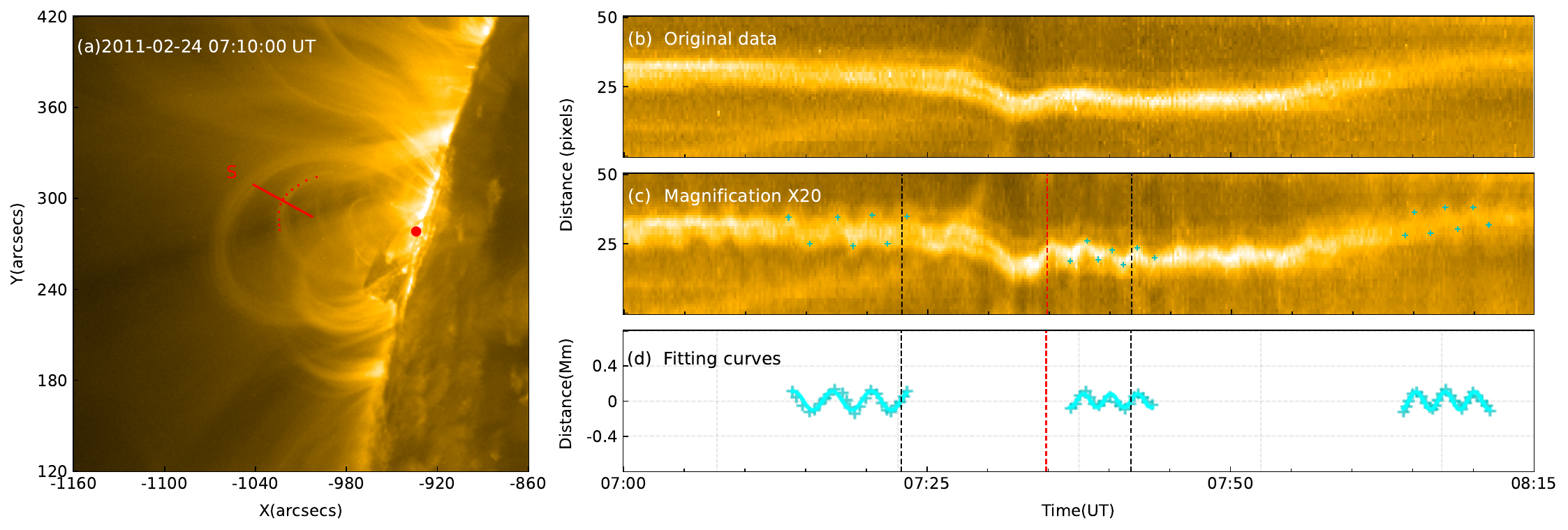} 
\caption{M3.5-class flare event on February 24, 2011. Target observation region: [-1160, -860, 120, 420] (arcsec). Other symbols are the same as in  Fig.\ref{fig:2014-01-28-C8.5}.}
\label{fig:2011-02-24-M3.5}
\end{figure*}

\begin{figure*}[ht!]
\centering
\includegraphics[width=\textwidth]{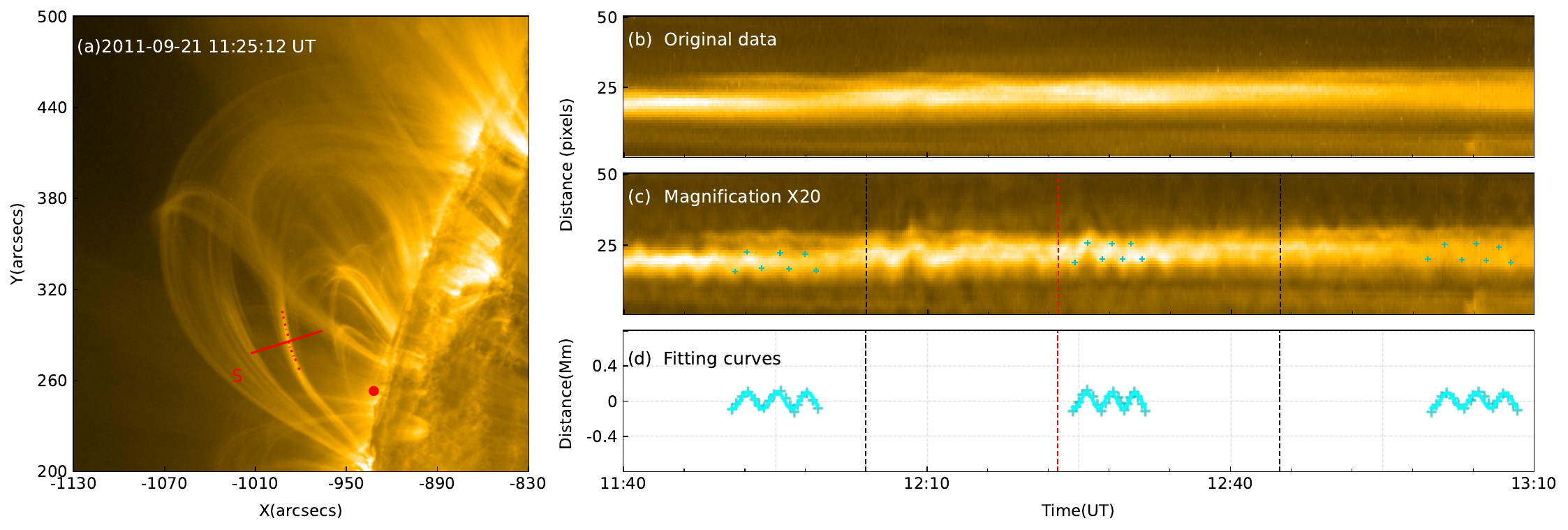} 
\caption{M1.8-class flare event on September 21, 2011. Target observation region: [-1130, -830, 200, 500] (arcsec). Other symbols are the same as in  Fig.\ref{fig:2014-01-28-C8.5}.}
\label{fig:2011-09-21-M1.8}
\end{figure*}

Figures~\ref{fig:2014-01-28-C8.5}---\ref{fig:2011-09-21-M1.8} show examples of decayless kink oscillations detected during periods of time that include solar flares. 

In the event presented in Figure~\ref{fig:2014-01-28-C8.5}, the raw time--distance map shows the appearance of the oscillation about the flare peak. The application of FA-VMM reveals the oscillations before and after the flare. During the flare, the amplitude increases by about 30\% compared to the pre-flare level, from  $0.070\pm0.004$~Mm to $0.091\pm0.009$~Mm. 

Figure~\ref{fig:2011-02-24-M3.5} presents an event in which the oscillation amplitude decreases during the flare by about 30\%, from 
$0.067 \pm 0.007$~Mm to $0.047 \pm 0.005$~Mm. After the flare, the amplitude recovers to $0.062 \pm 0.004$~Mm.

In Figure~\ref{fig:2011-09-21-M1.8} we show an event in which the amplitude is not affected by the flare, and is $0.058\pm0.006$~Mm before the flare, $0.057 \pm 0.008$~Mm during the flare, and $0.054 \pm 0.006$~Mm after the flare.

\FloatBarrier 
\FloatBarrier

\FloatBarrier 
\FloatBarrier 

\FloatBarrier 
\FloatBarrier 

\subsection{Parameters of Decayless Kink Oscillations}

\begin{figure*}[ht!]
\centering
\includegraphics[width=\textwidth]{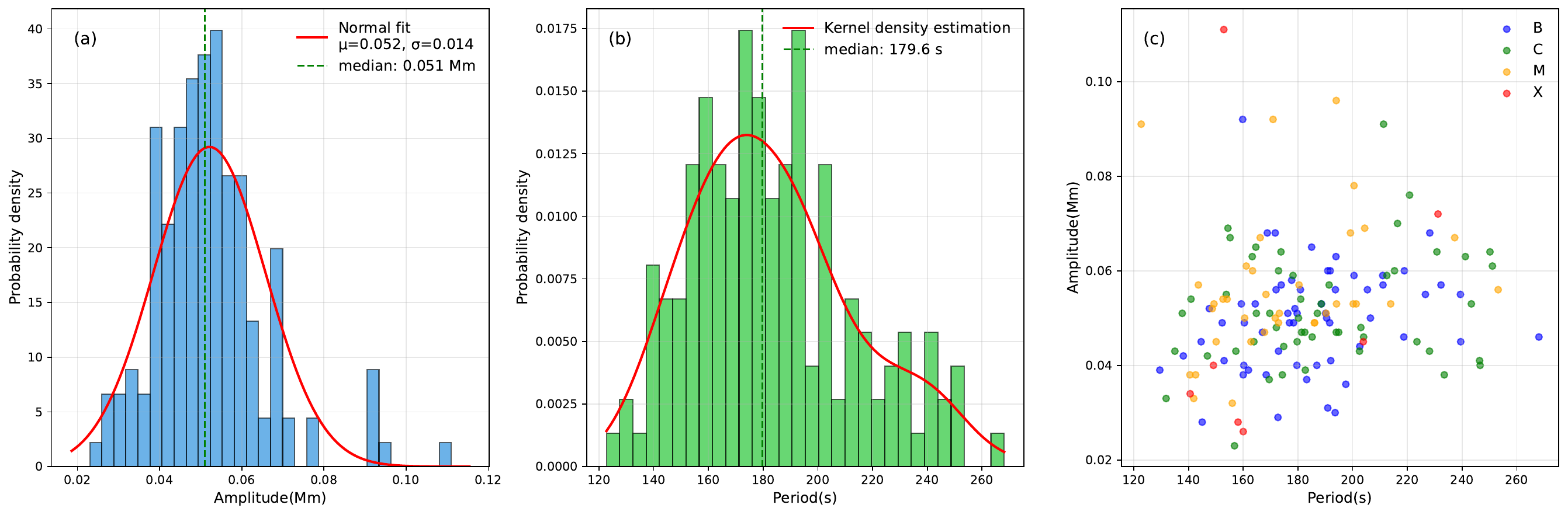} 
\caption{Distribution of oscillation parameters for all samples (during-flare). (a) Amplitude distribution histogram, (b) Period distribution histogram, (c) Amplitude--period scatter plot, color-coded by flare class.}
\label{fig:Statistical_distribution}
\end{figure*}

Figure~\ref{fig:Statistical_distribution} shows the statistical properties of the detected oscillations. The amplitudes range from 0.023~Mm to 0.111~Mm and exhibit an approximately normal distribution. The mean amplitude is $0.052 \pm 0.014$~Mm (with errors estimated as the standard deviation), and the median is 0.051~Mm. Approximately 68\% of the events have amplitudes concentrated in the 0.03--0.07~Mm range. These amplitude values are lower than the true values due to projection effects. The difference between the measured and true amplitudes depends on the angle between the polarization plane and the line of sight, which varies from case to case.

The oscillation periods range from 122~s to 268~s, the majority between 150~s and 220~s. The mean period is $183 \pm 29$~s, with a median of 179.6~s. The period distribution is broader than that of the amplitude, with a coefficient of variation (standard deviation divided by the mean) of 16.5\%. 
The projected amplitude does not show a strong correlation with the oscillation period (Pearson correlation coefficient $r \approx 0.16$, $p = 0.045$). No significant differences in the amplitude--period scaling are observed for flares of different classes.

\FloatBarrier 
\FloatBarrier

\subsection{Evolution of the amplitude}

\FloatBarrier 
\begin{figure*}[ht!]
\centering
\includegraphics[width=\textwidth]{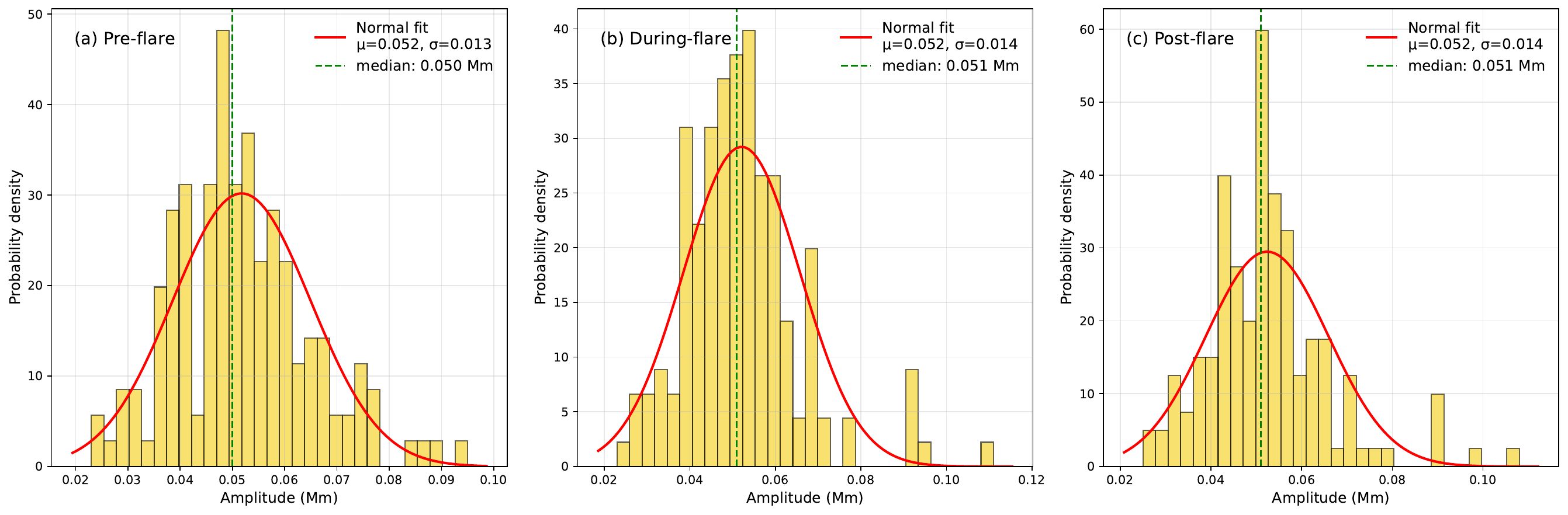} 
\caption{Amplitude distributions for pre-flare, during-flare, and post-flare phases.}
\label{fig:Amplitude_distribution_three_stages}
\end{figure*}
\FloatBarrier

Figure~\ref{fig:Amplitude_distribution_three_stages} shows distributions of the oscillation amplitudes before, during and after the flare. In the pre-flare phase, the amplitudes range from 0.023 Mm to 0.095 Mm, with the mean of $0.052 \pm 0.013$~Mm and a median of 0.050~Mm. During and after flares, amplitudes remain practically in the same range, from 0.023 Mm to 0.111 Mm, with the mean of $0.052 \pm 0.014$~Mm and the median of 0.051 Mm, and from 0.025~Mm to 0.108~Mm, with the mean of $0.052 \pm 0.014$~Mm and a median of 0.051~Mm, respectively. 

Despite the statistical characteristics of the decayless kink oscillation amplitude being not affected by flares, However, Section~\ref{sec:exa} evidences that in specific events the amplitude may evolve with the flare phase.

\begin{figure*}[ht!]
\centering
\includegraphics[width=\textwidth]{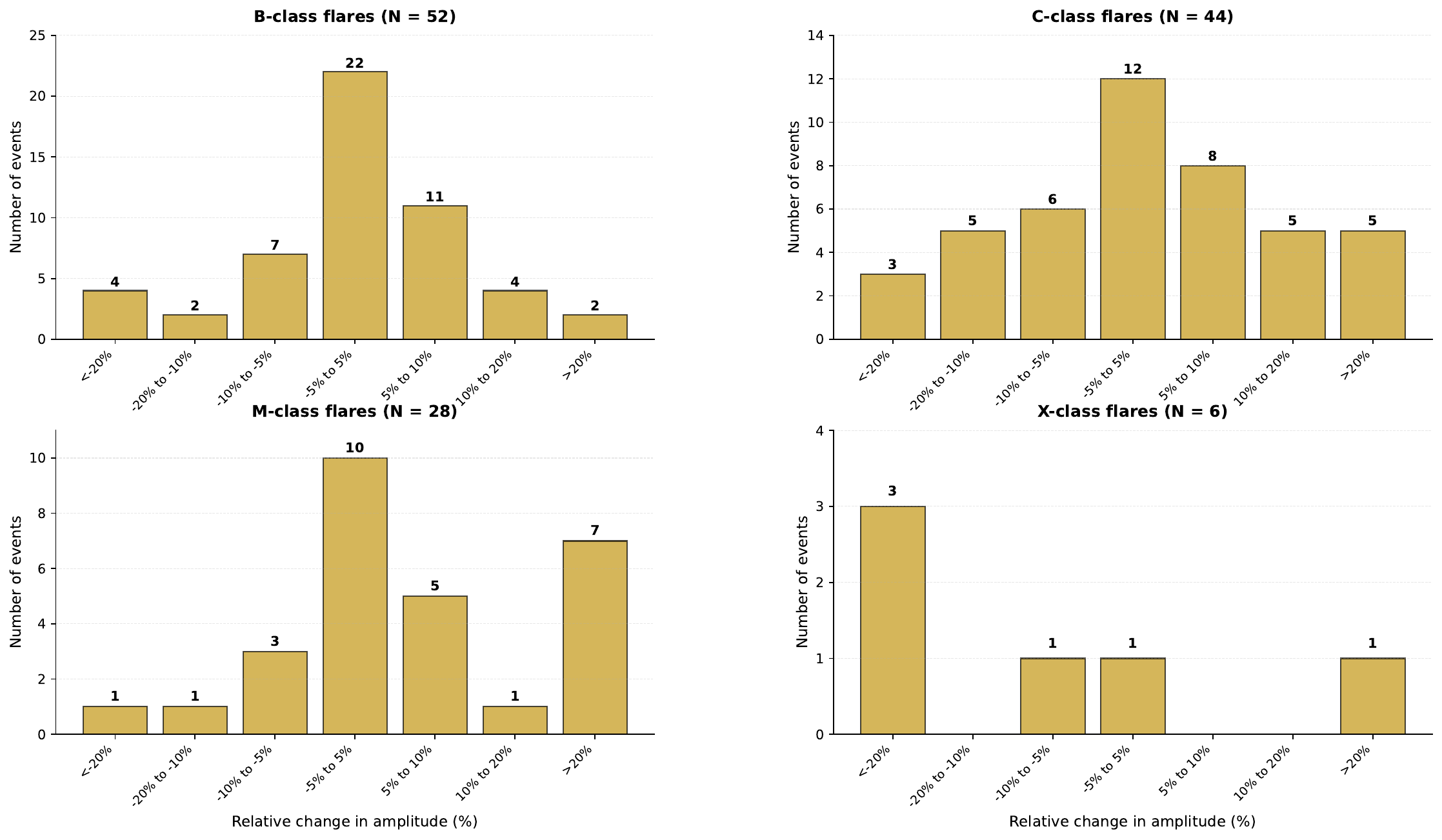}
\caption{Relative changes in the amplitudes of decayless kink oscillations during flares in comparison with the pre-flare value.}
\label{fig:Amplitude_change_histogram_1}
\end{figure*}

Figure~\ref{fig:Amplitude_change_histogram_1} shows the histogram of relative changes in the amplitude during the flare compared to the pre-flare value. The relative change is expressed as a percentage and calculated as $(A_{\text{during}} - A_{\text{pre}})/A_{\text{pre}}$, where $A_{\text{during}}$ and $A_{\text{pre}}$ denote the amplitudes during and before the flare, respectively. For each flare class, we determine the number of events in which the amplitude change falls within one of the following ranges: a decrease or increase of more than 20\%, from 10\% to 20\%, from 5\% to 10\%, and within $\pm 5$\% of the pre-flare value.
It is evident that in the majority of cases this relative change is negligible, less than 5\%.  In 104 out of 130 analysed events (i.e. 80\%), the relative change in amplitude is below 20\%.
Changes not exceeding 20\% are observed in 88.5\% of B-class flares, 81.8\% of C-class flares, 71.4\% of M-class flares, and 33.3\% of X-class flares. The latter value may not be representative due to the small number of events (only six). 
The increases and decreases in amplitude appear to be practically equally likely.

\begin{figure*}[ht!]
\centering
\includegraphics[width=\textwidth]{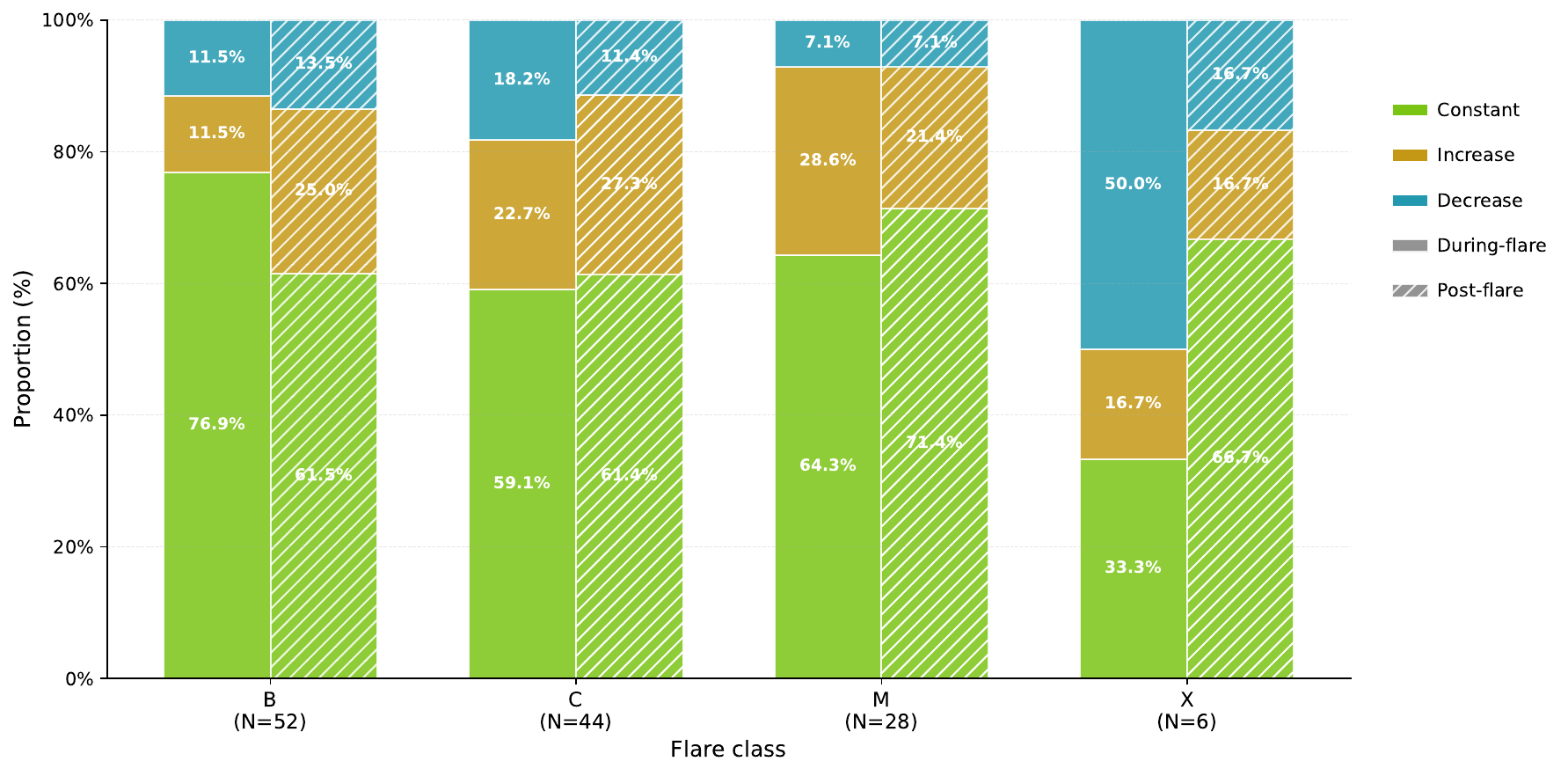}
\caption{Relative changes in the kink oscillation amplitude during and after a flare in comparison with the pre-flare value for the B, C, M, and X-class flares. The increase or decrease corresponds to the change by more than 10\%. The number of analysed events, $N$, is shown in the bottom. }
\label{fig:Amplitude_change_histogram_2}
\end{figure*}

Figure~\ref{fig:Amplitude_change_histogram_2} compares increases and decreases in the amplitude by more than 10\% during and after the flare relative to the pre-flare amplitude, separately for different flare classes. The vertical bars show the fractions (in percent) of amplitude increases, decreases, and cases of (nearly) constant amplitude, defined as changes of less than 10\%. The hatched bars show the same but for the post-flare changes.
The fractions of events with modest amplitude changes during the flare are 23\% (B), 41\% (C), 36\% (M), and 67\% (X). For post-flare amplitudes, the corresponding fractions are 39\% (B), 39\% (C), 29\% (M), and 33\% (X). Similarly to Figure~\ref{fig:Amplitude_change_histogram_1}, the results for X-class flares may be affected by the small number of events.

\FloatBarrier 
\FloatBarrier

\FloatBarrier 
\FloatBarrier

\begin{figure*}[ht!]
\centering
\includegraphics[width=\textwidth]{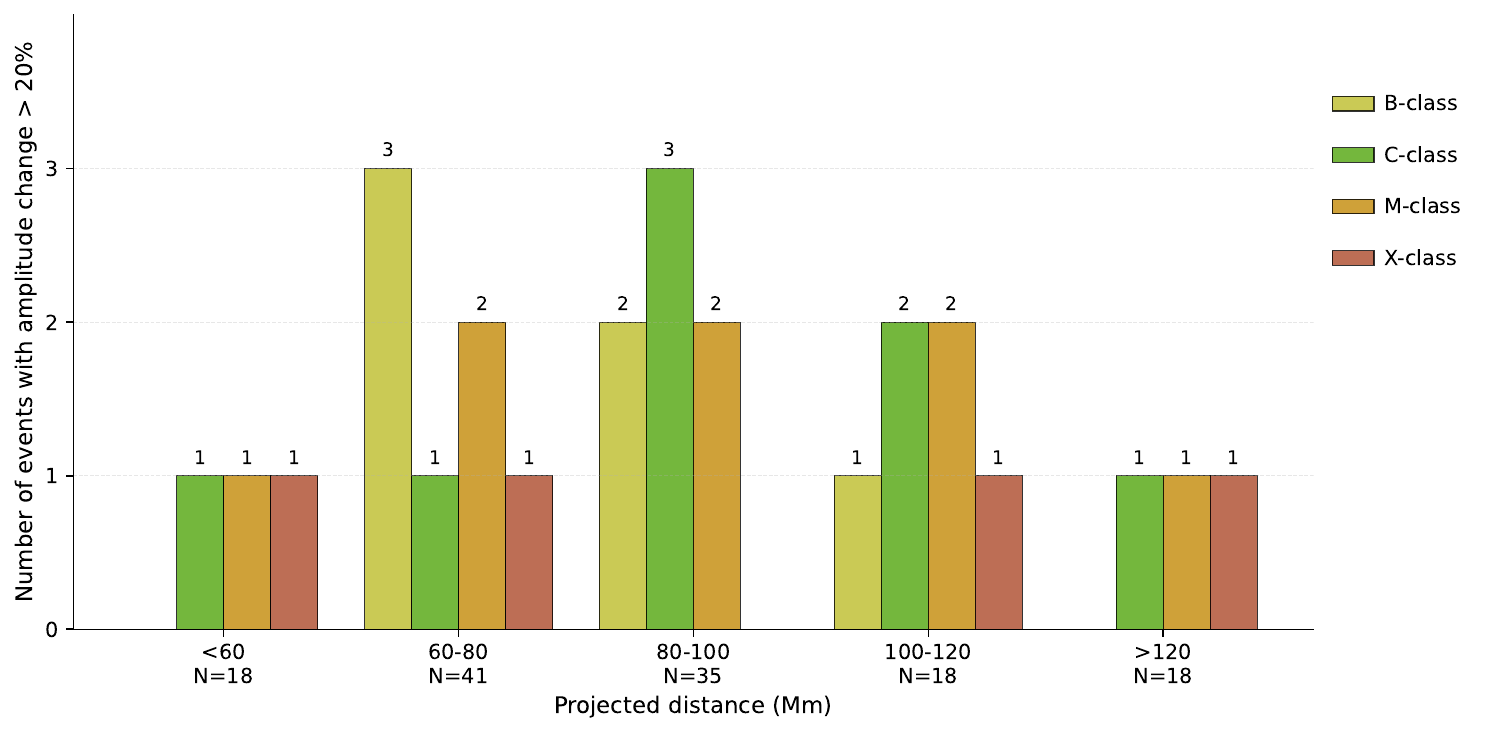}
\caption{Relationship between the loop location and modest amplitude change. The horizontal shows the projected distance from the flare centre to the oscillating loop. The vertical axis shows the number of events with amplitude change exceeding $20\%$. Different colours represent different flare classes (B, C, M, X). The value $N$ below each distance bin indicates the total number of events in that bin.}
\label{fig:Amplitude_change_histogram_3}
\end{figure*}

Figure~\ref{fig:Amplitude_change_histogram_3} shows the distribution of the events in which the amplitude changes by more than 20\% with the distance between the centre of the oscillating loop and the flare site, separately for different flare classes. 
The events were divided into five groups based on projected distance: $<$60~Mm, 60–-80~Mm, 80–-100~Mm, 100–-120~Mm, and $>$120~Mm. The numbers of events with modest amplitude changes ($>20\%$) in each group are 3, 7, 7, 6, and 3, corresponding to 16.7\%, 17.1\%, 20.0\%, 33.3\%, and 16.7\% of the total number of detected events in each group.

\FloatBarrier 
\FloatBarrier

\section{Discussion and Conclusions}
\label{sec:con}
This study analysed 130 SXR solar flare events that lack quasi-periodic pulsations in soft X-rays and show no visible oscillations in SDO/AIA 171~Å images. In the vicinity of each flare epicentre, we selected a well-defined coronal loop. The imaging data cubes containing the loops, covering the time intervals before, during, and after each flare, were processed using the Fractional Anisotropy-based Video Motion Magnification (FA-VMM) technique. This allowed us to detect low-amplitude decayless periodic transverse displacements of the loops, known as decayless kink oscillations. Our main findings are as follows.

Each of the 130 analysed loops is found to undergo decayless kink oscillations before, during, and after the flare. This confirms the ubiquitous and persistent nature of decayless kink oscillations of coronal loops. 

The oscillation periods range from 122~s to 268~s, which is consistent with the range reported in previous statistical studies of this phenomenon \citep[e.g.,][]{2015A&A...583A.136A}. The projected amplitudes range from 0.023~Mm to 0.111~Mm, which is at the lower end of the previously detected amplitudes. This discrepancy may be explained by a selection effect: the events analysed here were specifically chosen to show no kink oscillations visible without motion magnification. The amplitude does not correlate with the oscillation period, which is also consistent with previous findings \citep{2016A&A...591L...5N}. This result supports the conclusion that the energy supply mechanism is not based on the leakage of low-atmospheric oscillations into the corona.

For each oscillating loop, we estimated the oscillation amplitude before, during, and after the flare. The evolution of the oscillation period was not addressed in this study. This was motivated by the absence of any modification of the decayless kink oscillation period by a flare, as established in previous case studies \citep{2013A&A...552A..57N, 2021A&A...652L...3M, 2022RAA....22j5017S}, and corroborated by a visual inspection of our dataset. Thus, here we focus on amplitude evolution, which is directly relevant to the energy supply question. 

Across all analyzed time intervals, the average amplitude was found to be the same for all flare classes. This finding is consistent with the results of \citet{2022RAA....22j5017S}, who found no significant changes in the parameters of decayless kink oscillations in an active region during a C2.0 flare. On the other hand, our finding highlights the uniqueness of the event reported by \citet{2021A&A...652L...3M}, in which the decayless kink amplitude increased during and after a flare by up to an order of magnitude.

However, in individual loops, the amplitude may change.
For B-, C-, and M-class flares, the fraction of events exhibiting a modest change in amplitude (greater than 10\%) is approximately 23\%, 41\%, and 36\%, respectively. In B- and C-class flares, amplitude increases and decreases are almost equally likely, whereas for M-class flares, amplitude increases occur four times more often than decreases. In X-class flares, the amplitude changes in two-thirds of cases, with decreases occurring three times more frequently than increases. Nevertheless, this estimate may not be reliable due to the small number of events, only six. 

The likelihood of a modest amplitude change ($>20\%$) shows weak dependence on the projected distance between the flare site and the centre of the oscillating loop. For projected distances of $<60$~Mm, 60–-80~Mm, 80-–100~Mm, 100–-120~Mm, and $>120$~Mm, the fractions of events with modest amplitude changes are 16.7\%, 17.1\%, 20.0\%, 33.3\%, and 16.7\%, respectively. 

Thus, our results show that the amplitude of decayless kink oscillations does not experience a major change in response to a flare occurring nearby, at least for less powerful flare classes. This finding suggests that the dynamic processes accompanying a flare, such as the coronal blast wave and reconnection inflows, have little effect on the energy supply to oscillating loops. The energy supply mechanism is likely based on low-atmospheric dynamics, which is less influenced by a flare. 

\begin{acknowledgments}
We thank the SDO/AIA science team for providing the data, and the developers of the fractional anisotropy-based video magnification technique. 
V.M.N. is supported by ERC grant 101201424 (ACDCSUN) and the BK21 FOUR program through the National Research Foundation of Korea (NRF) under the Ministry of Education (MoE) (Kyung Hee University, Human Education Team for the Next Generation of Space Exploration), and the Global-Learning \& Academic research institution for Master’s/PhD students, the Postdocs (G-LAMP) Program of the NRF grant funded by the MoE (RS-2025-25442355).
D.Y. was supported by the National Natural Science Foundation of China (NSFC: 12473050), the Guangdong Natural Science Funds for Distinguished Young Scholars (2023B1515020049), the Shenzhen Science and Technology Project (JCYJ20240813104805008), and the Specialized Research Fund for State Key Laboratory of Solar Activity and Space Weather.
\end{acknowledgments}

\section*{Appendix}
This appendix provides the full list of parameters for all 160 events analyzed in this study. 
The table includes event time, flare class, flare start-peak-end times (UT), target observation region (arcsec), 
{the coordinates of flare centre and slice centre (arcsec)}, the projected distance between them (Mm), 
pre-flare, during-flare, and post-flare amplitudes (Mm), and the mean period (s) calculated from the three phases. 
Uncertainties are $1\sigma$ from sine fitting. Ellipsis (\dots) indicates no data available.
\FloatBarrier 
\setlength{\tabcolsep}{2pt}
\scriptsize
\begin{sidewaystable}
\begin{longtable}{l l l l l l l l l l l}
\caption{Parameters of the Analyzed Events \label{tab:data}} \\
\toprule
Event & Flare class & Flare time (UT) & Observation region & Flare centre & Slice centre & Proj. dist. & $A_{\rm pre}$ & $A_{\rm during}$ & $A_{\rm post}$ & $P_{\rm mean}$ \\
& & & (arcsec) & (arcsec) & (arcsec) & (Mm) & (Mm) & (Mm) & (Mm) & (s) \\
\midrule
\endfirsthead
\bottomrule
\endlastfoot
2016-12-07 & B3.6 & 17:52-18:09-18:20 & [660, 960, -270, 30] & (789.8, -106.3) & (803.2, -31.4) & 55.30 & 0.048$\pm$0.003 & 0.049$\pm$0.004 & 0.042$\pm$0.005 & 172.7$\pm$1.2 \\
2016-09-27 & B3.6 & 05:36-05:44-05:58 & [500, 800, -480, -180] & (594.7, -332.9) & (628.3, -446.0) & 85.74 & 0.075$\pm$0.005 & 0.047$\pm$0.005 & 0.047$\pm$0.004 & 157.4$\pm$1.1 \\
2016-12-01 & B3.7 & 11:40-11:47-11:59 & [-620, -320, -270, 30] & (-514.3, -127.1) & (-520.7, -218.3) & 66.43 & 0.063$\pm$0.005 & 0.065$\pm$0.006 & 0.089$\pm$0.008 & 186.7$\pm$1.2 \\
2016-09-17 & B4.0 & 05:55-06:05-06:18 & [-950, -650, 90, 390] & (-775.3, 219.5) & (-794.0, 303.5) & 62.52 & 0.048$\pm$0.003 & 0.051$\pm$0.002 & 0.045$\pm$0.004 & 177.1$\pm$1.0 \\
2016-04-21 & B4.7 & 04:58-05:10-05:23 & [870, 1170, -50, 250] & (958.6, 107.2) & (1004.5, -12.6) & 93.22 & 0.061$\pm$0.002 & 0.060$\pm$0.004 & 0.058$\pm$0.005 & 187.0$\pm$1.0 \\
2016-04-14-2 & B4.8 & 14:37-14:59-15:10 & [-10, 290, 30, 330] & (70.4, 248.7) & (212.2, 122.7) & 137.80 & 0.055$\pm$0.005 & 0.057$\pm$0.003 & 0.045$\pm$0.006 & 229.0$\pm$1.7 \\
2016-04-14-1 & B4.8 & 14:37-14:59-15:10 & [-70, 230, 250, 550] & (77.8, 297.2) & (48.2, 416.6) & 89.31 & 0.053$\pm$0.003 & 0.053$\pm$0.003 & 0.055$\pm$0.003 & 192.6$\pm$0.9 \\
2016-06-25 & B4.9 & 11:01-11:11-11:25 & [-1170, -870, 40, 340] & (-906.8, 213.5) & (-1004.1, 124.1) & 96.02 & 0.046$\pm$0.004 & 0.049$\pm$0.004 & 0.050$\pm$0.005 & 173.9$\pm$1.2 \\
2016-09-25 & B5.1 & 23:25-23:37-23:49 & [260, 560, -510, -210] & (364.4, -340.5) & (318.9, -283.9) & 52.72 & 0.038$\pm$0.004 & 0.045$\pm$0.003 & 0.051$\pm$0.004 & 146.6$\pm$0.9 \\
2016-09-19 & B5.3 & 07:32-07:42-07:52 & [-630, -330, 60, 360] & (-385.3, 127.5) & (-456.2, 289.1) & 128.17 & 0.040$\pm$0.002 & 0.040$\pm$0.003 & 0.048$\pm$0.004 & 194.2$\pm$1.1 \\
2015-07-21 & B5.3 & 05:23-05:47-06:20 & [-200, 100, 80, 380] & (-47.7, 192.2) & (-86.3, 129.8) & 53.34 & 0.041$\pm$0.003 & 0.039$\pm$0.004 & 0.038$\pm$0.004 & 131.8$\pm$1.0 \\
2015-08-03 & B5.4 & 05:37-05:48-05:56 & [100, 400, -150, 150] & (140.6, -6.8) & (227.6, -2.2) & 63.31 & 0.052$\pm$0.003 & 0.055$\pm$0.003 & 0.058$\pm$0.004 & 223.8$\pm$1.2 \\
2015-04-09 & B5.7 & 00:33-00:39-00:51 & [70, 370, -290, 10] & (136.5, -119.5) & (289.9, -143.7) & 112.82 & 0.038$\pm$0.002 & 0.030$\pm$0.003 & 0.025$\pm$0.003 & 183.6$\pm$1.3 \\
2014-04-06 & B6.1 & 15:40-16:04-16:21 & [160, 460, 70, 370] & (325.4, 149.6) & (367.4, 95.0) & 50.04 & 0.059$\pm$0.003 & 0.051$\pm$0.003 & \dots & 197.0$\pm$1.4 \\
2016-10-04 & B6.1 & 03:53-04:09-04:24 & [-770, -470, -30, 270] & (-646.3, 127.5) & (-687.0, 36.8) & 72.23 & 0.075$\pm$0.005 & 0.068$\pm$0.003 & 0.071$\pm$0.005 & 216.8$\pm$1.0 \\
2015-07-20 & B6.4 & 04:32-04:52-05:05 & [0, 300, 160, 460] & (172.0, 198.4) & (195.5, 317.4) & 88.16 & 0.038$\pm$0.002 & 0.037$\pm$0.004 & 0.038$\pm$0.002 & 177.2$\pm$1.3 \\
2015-07-04 & B6.4 & 20:33-20:45-20:55 & [-870, -570, -440, -140] & (-702.2, -304.3) & (-814.1, -186.5) & 118.05 & 0.050$\pm$0.004 & 0.052$\pm$0.003 & 0.053$\pm$0.004 & 163.1$\pm$0.9 \\
2016-09-03 & B6.4 & 00:42-01:03-01:13 & [-810, -510, -100, 200] & (-762.0, 73.0) & (-708.9, -72.8) & 112.69 & 0.057$\pm$0.002 & 0.057$\pm$0.00 & 0.052$\pm$0.004 & 175.1$\pm$1.0 \\
2010-08-04 & B6.5 & 09:13-09:20-09:39 & [-1160, -860, 100, 400] & (-911.1, 278.9) & (-1029.2, 272.0) & 85.93 & 0.061$\pm$0.005 & 0.058$\pm$0.004 & 0.060$\pm$0.004 & 178.2$\pm$1.0 \\
2017-03-29 & B6.5 & 23:19-23:32-23:50 & [0, 300, 220, 520] & (120.7, 348.3) & (148.8, 459.9) & 83.66 & 0.050$\pm$0.003 & 0.057$\pm$0.003 & 0.057$\pm$0.004 & 210.8$\pm$1.0 \\
2015-07-05 & B6.7 & 06:52-07:00-07:08 & [-840, -530, -440, -140] & (-636.4, -312.0) & (-750.0, -259.8) & 88.38 & 0.051$\pm$0.003 & 0.053$\pm$0.003 & 0.058$\pm$0.002 & 166.2$\pm$0.8 \\
2012-04-07 & B6.8 & 19:08-19:55-20:16 & [670, 970, 210, 510] & (796.8, 291.1) & (860.3, 363.7) & 70.08 & \dots & 0.038$\pm$0.003 & 0.041$\pm$0.003 & 162.6$\pm$1.2 \\
2012-08-15 & B6.8 & 14:48-14:54-15:12 & [470, 770, -420, -120] & (619.3, -259.8) & (682.6, -260.8) & 45.95 & 0.061$\pm$0.003 & 0.063$\pm$0.002 & 0.063$\pm$0.004 & 188.0$\pm$0.8 \\
2016-02-29 & B6.9 & 04:19-04:30-04:41 & [0, 300, -100, 200] & (144.1, 12.2) & (202.2, 101.2) & 77.15 & 0.048$\pm$0.004 & 0.050$\pm$0.004 & 0.049$\pm$0.003 & 203.9$\pm$1.1 \\
2012-04-05 & B6.9 & 10:11-10:30-10:49 & [290, 590, 270, 570] & (404.5, 316.8) & (474.9, 372.9) & 65.39 & 0.053$\pm$0.005 & 0.056$\pm$0.006 & 0.054$\pm$0.003 & 198.8$\pm$1.5 \\
2016-04-16 & B7.0 & 05:06-05:13-05:38 & [300, 600, 80, 380] & (425.1, 236.2) & (367.5, 135.4) & 84.32 & 0.035$\pm$0.004 & 0.031$\pm$0.002 & 0.034$\pm$0.002 & 193.0$\pm$1.0 \\
2015-02-21 & B7.1 & 18:20-18:50-19:41 & [-410, -110, 320, 620] & (-258.4, 436.9) & (-323.7, 531.3) & 83.33 & 0.063$\pm$0.003 & 0.050$\pm$0.005 & 0.059$\pm$0.003 & 181.8$\pm$1.1 \\
2010-07-31 & B7.4 & 18:32-19:09-19:26 & [-800, -500, 30, 330] & (-599.3, 226.7) & (-693.4, 156.0) & 85.49 & 0.060$\pm$0.003 & 0.059$\pm$0.006 & 0.063$\pm$0.004 & 198.6$\pm$1.2 \\
2016-04-15-2 & B7.6 & 02:46-02:50-03:03 & [90, 390, 30, 330] & (231.4, 223.5) & (344.3, 110.1) & 116.24 & 0.059$\pm$0.005 & 0.060$\pm$0.005 & \dots & 217.4$\pm$1.7 \\
2016-04-15-1 & B7.6 & 02:46-02:50-03:03 & [30, 330, 260, 560] & (147.4, 281.6) & (183.8, 428.6) & 109.99 & 0.038$\pm$0.004 & 0.041$\pm$0.003 & 0.047$\pm$0.003 & 194.0$\pm$1.2 \\
2011-02-09 & B7.9 & 22:54-23:03-23:13 & [820, 1120, 100, 400] & (927.1, 295.5) & (1018.7, 325.4) & 70.03 & 0.056$\pm$0.005 & 0.059$\pm$0.005 & 0.057$\pm$0.006 & 204.7$\pm$1.9 \\
2011-02-28 & B7.9 & 04:17-04:28-04:34 & [-900, -600, 360, 660] & (-657.1, 515.2) & (-736.3, 566.3) & 68.55 & 0.060$\pm$0.006 & 0.056$\pm$0.005 & 0.058$\pm$0.004 & 203.9$\pm$1.4 \\
2015-08-01 & B8.0 & 01:14-01:31-01:37 & [-760, -460, 80, 380] & (-538.8, 230.3) & (-636.4, 244.6) & 71.65 & 0.049$\pm$0.002 & 0.056$\pm$0.003 & 0.051$\pm$0.003 & 177.7$\pm$0.8 \\
2010-07-27 & B8.0 & 03:48-03:57-04:09 & [200, 500, -620, -320] & (330.8, -444.3) & (328.8, -343.7) & 73.05 & 0.064$\pm$0.004 & 0.044$\pm$0.002 & 0.043$\pm$0.003 & 200.4$\pm$1.0 \\
2014-04-12 & B8.1 & 15:39-16:20-16:34 & [-1100, -800, -380, -80] & (-853.8, -272.9) & (-937.6, -319.2) & 69.48 & 0.046$\pm$0.003 & 0.045$\pm$0.004 & \dots & 235.1$\pm$2.0 \\
\end{longtable}
\end{sidewaystable}

\begin{sidewaystable}
\begin{longtable}{l l l l l l l l l l l}
\toprule
Event & Flare class & Flare time (UT) & Observation region & Flare centre & Slice centre & Proj. dist. & $A_{\rm pre}$ & $A_{\rm during}$ & $A_{\rm post}$ & $P_{\rm mean}$ \\
& & & (arcsec) & (arcsec) & (arcsec) & (Mm) & (Mm) & (Mm) & (Mm) & (s) \\
\midrule
\endfirsthead
\bottomrule
\endlastfoot
2013-02-24 & B8.1 & 07:10-07:19-07:32 & [900, 1200, -20, 280] & (965.1, 148.3) & (1053.5, 34.7) & 109.90 & \dots & 0.046$\pm$0.009 & 0.052$\pm$0.003 & 259.4$\pm$3.8 \\
2011-04-23 & B8.1 & 12:43-12:52-13:07 & [650, 950, 220, 520] & (715.6, 357.4) & (733.6, 421.4) & 48.29 & 0.047$\pm$0.003 & \dots & 0.032$\pm$0.004 & 191.5$\pm$1.8 \\
2011-04-28 & B8.3 & 10:02-10:19-10:28 & [420, 720, 240, 540] & (592.3, 390.3) & (641.2, 433.3) & 47.30 & 0.075$\pm$0.004 & 0.068$\pm$0.006 & 0.075$\pm$0.006 & 165.7$\pm$1.0 \\
2016-07-17 & B8.4 & 18:59-19:15-19:26 & [-120, 180, -190, 110] & (46.6, -16.2) & (51.8, -114.3) & 71.34 & 0.046$\pm$0.004 & 0.049$\pm$0.002 & 0.051$\pm$0.005 & 179.9$\pm$1.2 \\
2016-04-20 & B8.5 & 02:24-02:36-02:46 & [900, 1200, -30, 270] & (912.7, 199.2) & (1020.7, 31.1) & 145.17 & 0.031$\pm$0.005 & 0.028$\pm$0.003 & 0.031$\pm$0.005 & 146.9$\pm$1.8 \\
2016-02-06 & B8.5 & 03:56-04:21-04:39 & [-100, 200, -240, 60] & (83.8, -104.7) & (61.1, -213.4) & 80.68 & 0.038$\pm$0.005 & 0.036$\pm$0.002 & 0.038$\pm$0.003 & 195.3$\pm$1.4 \\
2015-05-17 & B8.6 & 02:08-02:15-02:24 & [370, 670, 200, 500] & (428.4, 320.7) & (444.7, 395.7) & 55.83 & 0.045$\pm$0.005 & 0.046$\pm$0.003 & 0.043$\pm$0.002 & 214.0$\pm$1.5 \\
2016-04-13 & B8.6 & 07:35-07:57-08:16 & [-350, -50, 120, 420] & (-256.8, 288.8) & (-217.2, 387.1) & 76.98 & 0.043$\pm$0.003 & 0.042$\pm$0.003 & 0.044$\pm$0.004 & 134.0$\pm$0.8 \\
2012-06-28 & B8.8 & 09:09-09:23-09:27 & [-270, 30, -440, -140] & (-121.7, -283.8) & (-119.2, -216.1) & 49.20 & 0.049$\pm$0.003 & 0.049$\pm$0.004 & 0.045$\pm$0.007 & 208.2$\pm$2.4 \\
2012-06-27 & B8.9 & 01:40-01:50-02:00 & [-1180, -880, 120, 420] & (-884.0, 202.1) & (-1006.2, 244.4) & 93.98 & 0.045$\pm$0.004 & 0.041$\pm$0.003 & 0.039$\pm$0.005 & 146.9$\pm$1.1 \\
2016-02-20 & B8.9 & 00:49-01:31-01:42 & [-1110, -810, -560, -260] & (-868.3, -457.4) & (-958.2, -355.1) & 98.93 & 0.045$\pm$0.002 & 0.049$\pm$0.003 & 0.043$\pm$0.003 & 156.9$\pm$0.8 \\
2011-04-16 & B9.0 & 20:52-21:02-21:26 & [-670, -370, 160, 460] & (-499.0, 305.3) & (-613.4, 245.6) & 93.75 & 0.058$\pm$0.006 & 0.055$\pm$0.004 & 0.063$\pm$0.005 & 212.5$\pm$1.5 \\
2014-04-08 & B9.1 & 18:36-18:51-19:21 & [500, 800, -240, 60] & (633.1, -124.9) & (560.8, -192.7) & 71.97 & 0.044$\pm$0.004 & 0.056$\pm$0.004 & 0.060$\pm$0.003 & 170.9$\pm$0.9 \\
2014-05-20 & B9.2 & 16:42-16:51-17:01 & [-1100, -800, -430, -130] & (-868.9, -186.1) & (-951.0, -329.6) & 120.09 & 0.059$\pm$0.006 & 0.053$\pm$0.003 & 0.043$\pm$0.005 & 176.7$\pm$1.3 \\
2016-06-13 & B9.2 & 04:35-04:49-05:02 & [870, 1170, 30, 330] & (940.0, 181.7) & (986.9, 66.6) & 90.31 & 0.042$\pm$0.004 & 0.040$\pm$0.003 & 0.040$\pm$0.003 & 183.3$\pm$1.2 \\
2014-01-24 & B9.2 & 09:58-10:37-10:41 & [-600, -300, -100, 200] & (-446.5, -45.3) & (-532.5, -21.4) & 64.83 & 0.040$\pm$0.003 & 0.043$\pm$0.003 & 0.064$\pm$0.003 & 182.1$\pm$0.9 \\
2016-04-11-2 & B9.3 & 21:59-22:12-22:19 & [-650, -350, 230, 530] & (-562.0, 259.2) & (-416.5, 320.5) & 114.69 & 0.039$\pm$0.003 & 0.039$\pm$0.003 & 0.043$\pm$0.002 & 164.9$\pm$1.0 \\
2016-04-11-1 & B9.3 & 21:59-22:12-22:19 & [-600, -300, 10, 310] & (-473.4, 196.1) & (-416.8, 117.8) & 70.18 & \dots & 0.038$\pm$0.003 & 0.037$\pm$0.003 & 164.7$\pm$1.2 \\
2014-05-16 & B9.4 & 08:57-09:03-09:14 & [-600, -300, 100, 400] & (-440.5, 210.8) & (-422.8, 296.0) & 63.25 & 0.050$\pm$0.004 & 0.051$\pm$0.003 & 0.046$\pm$0.003 & 205.2$\pm$1.2 \\
2016-01-27 & B9.4 & 05:26-05:37-05:52 & [-780, -480, 80, 380] & (-610.7, 226.1) & (-611.9, 110.2) & 84.17 & 0.038$\pm$0.003 & 0.040$\pm$0.006 & 0.050$\pm$0.004 & 164.0$\pm$1.3 \\
2015-02-13 & B9.6 & 02:20-02:28-02:51 & [-460, -160, 110, 410] & (-326.5, 241.7) & (-296.6, 159.8) & 63.41 & 0.051$\pm$0.003 & 0.060$\pm$0.003 & 0.053$\pm$0.003 & 178.8$\pm$0.8 \\
2015-02-28 & B9.7 & 21:12-21:42-22:01 & [250, 550, -290, 10] & (412.2, -114.0) & (458.9, -185.9) & 62.27 & 0.023$\pm$0.004 & 0.052$\pm$0.004 & 0.057$\pm$0.004 & 150.9$\pm$1.5 \\
2015-05-03 & B9.8 & 04:07-04:19-04:38 & [-860, -560, -350, -50] & (-712.2, -211.3) & (-774.9, -93.2) & 97.14 & 0.065$\pm$0.003 & 0.068$\pm$0.004 & 0.066$\pm$0.006 & 195.1$\pm$1.1 \\
2012-06-11 & B9.8 & 12:32-12:37-13:06 & [-550, -250, -550, -250] & (-331.0, -408.1) & (-275.4, -457.3) & 53.91 & 0.037$\pm$0.002 & 0.029$\pm$0.006 & \dots & 167.8$\pm$2.3 \\
2012-06-04 & B9.9 & 15:35-15:42-16:05 & [-550, -250, -450, -150] & (-422.2, -315.5) & (-360.7, -204.9) & 91.98 & 0.086$\pm$0.006 & 0.092$\pm$0.005 & 0.089$\pm$0.004 & 157.3$\pm$0.7 \\
2016-07-16-1 & C1.0 & 09:19-09:45-09:49 & [-560, -260, -80, 220] & (-390.2, 29.5) & (-505.3, -3.1) & 86.92 & 0.059$\pm$0.005 & 0.051$\pm$0.007 & 0.058$\pm$0.003 & 165.0$\pm$1.2 \\
2016-07-16-2 & C1.0 & 09:19-09:45-09:49 & [-370, -70, -160, 140] & (-309.9, 9.5) & (-261.2, 59.7) & 50.76 & 0.048$\pm$0.003 & 0.051$\pm$0.004 & 0.048$\pm$0.004 & 168.0$\pm$1.0 \\
2011-03-25 & C1.0 & 16:47-17:13-17:22 & [-750, -450, -170, 130] & (-475.2, -138.3) & (-641.4, -27.8) & 144.99 & 0.065$\pm$0.005 & 0.060$\pm$0.003 & 0.070$\pm$0.005 & 217.9$\pm$1.2 \\
2011-01-14 & C1.0 & 12:52-13:07-13:25 & [-1130, -830, 270, 570] & (-906.8, 409.4) & (-1059.8, 459.3) & 116.85 & 0.060$\pm$0.007 & 0.061$\pm$0.005 & 0.058$\pm$0.004 & 250.7$\pm$1.9 \\
2017-03-31 & C1.1 & 22:54-23:01-23:10 & [-200, 100, -260, 40] & (-48.2, -58.1) & (-71.7, 12.3) & 53.91 & 0.050$\pm$0.003 & 0.045$\pm$0.006 & 0.051$\pm$0.003 & 199.7$\pm$1.5 \\
2013-01-01 & C1.2 & 08:47-09:06-09:13 & [-180, 120, 400, 700] & (-11.7, 487.1) & (-101.9, 445.3) & 72.18 & 0.055$\pm$0.003 & 0.057$\pm$0.005 & 0.057$\pm$0.004 & 192.6$\pm$1.2 \\
2017-02-23 & C1.3 & 20:36-20:53-21:17 & [-550, -250, 350, 650] & (-402.4, 440.0) & (-422.9, 519.0) & 59.33 & 0.031$\pm$0.002 & 0.051$\pm$0.006 & 0.063$\pm$0.006 & 156.2$\pm$1.2 \\
2015-01-25 & C1.4 & 11:56-12:12-12:21 & [-1200, -900, -300, 0] & (-972.6, -102.0) & (-1073.9, -54.6) & 81.25 & 0.042$\pm$0.005 & 0.039$\pm$0.007 & 0.053$\pm$0.004 & 180.4$\pm$2.1 \\
2014-01-10 & C1.4 & 07:06-07:27-08:26 & [430, 730, 30, 330] & (547.9, 189.9) & (593.4, 61.0) & 99.33 & 0.046$\pm$0.004 & 0.043$\pm$0.003 & 0.052$\pm$0.004 & 158.8$\pm$0.9 \\
2012-01-23 & C1.4 & 19:40-20:11-20:26 & [420, 720, 360, 660] & (537.7, 371.2) & (481.9, 585.9) & 161.14 & 0.055$\pm$0.005 & 0.059$\pm$0.004 & 0.062$\pm$0.004 & 177.0$\pm$1.0 \\

\end{longtable}
\end{sidewaystable}

\begin{sidewaystable}
\begin{longtable}{l l l l l l l l l l l}
\toprule
Event & Flare class & Flare time (UT) & Observation region & Flare centre & Slice centre & Proj. dist. & $A_{\rm pre}$ & $A_{\rm during}$ & $A_{\rm post}$ & $P_{\rm mean}$ \\
& & & (arcsec) & (arcsec) & (arcsec) & (Mm) & (Mm) & (Mm) & (Mm) & (s) \\
\midrule
\endfirsthead
\bottomrule
\endlastfoot
2015-01-04 & C1.5 & 00:32-00:53-01:28 & [-370, -70, -100, 200] & (-201.2, -15.9) & (-275.3, 100.2) & 100.06 & 0.045$\pm$0.003 & 0.043$\pm$0.004 & 0.045$\pm$0.002 & 182.7$\pm$1.2 \\
2016-07-24 & C1.5 & 04:50-05:01-05:22 & [880, 1180, -30, 270] & (938.7, 91.6) & (992.8, 4.0) & 74.85 & 0.064$\pm$0.005 & 0.069$\pm$0.005 & 0.108$\pm$0.005 & 153.8$\pm$0.8 \\
2013-04-06 & C1.7 & 11:15-11:42-11:54 & [0, 300, -500, -200] & (83.8, -292.9) & (126.1, -372.2) & 65.27 & 0.037$\pm$0.003 & 0.037$\pm$0.004 & 0.042$\pm$0.004 & 172.8$\pm$1.1 \\
2016-06-12 & C1.8 & 20:30-20:40-20:52 & [880, 1180, 20, 320] & (918.1, 240.7) & (986.6, 164.1) & 74.61 & 0.055$\pm$0.005 & 0.051$\pm$0.006 & 0.054$\pm$0.004 & 185.1$\pm$1.4 \\
2016-04-12 & C1.8 & 07:40-07:51-08:10 & [-540, -240, 120, 420] & (-410.5, 194.4) & (-489.8, 142.3) & 68.94 & 0.045$\pm$0.003 & 0.048$\pm$0.003 & 0.047$\pm$0.002 & 170.2$\pm$0.9 \\
2016-04-14 & C1.8 & 13:21-13:32-13:44 & [-50, 250, 70, 370] & (38.6, 282.0) & (201.8, 121.0) & 166.62 & 0.076$\pm$0.005 & 0.064$\pm$0.005 & 0.050$\pm$0.006 & 229.4$\pm$1.6 \\
2017-01-21 & C1.8 & 13:19-13:26-13:30 & [-780, -480, 110, 410] & (-640.6, 246.9) & (-705.3, 313.9) & 67.65 & 0.064$\pm$0.004 & 0.063$\pm$0.006 & 0.042$\pm$0.004 & 236.1$\pm$1.7 \\
2015-01-26 & C1.8 & 06:08-06:27-06:49 & [-250, 50, 200, 500] & (-89.6, 389.0) & (-116.5, 321.9) & 42.14 & $\dots$ & 0.040$\pm$0.003 & 0.042$\pm$0.004 & 246.0$\pm$2.3 \\
2015-01-27 & C1.9 & 09:11-09:24-09:40 & [-1200, -900, -60, 240] & (-978.7, 112.0) & (-1107.9, 148.0) & 97.42 & 0.030$\pm$0.003 & 0.023$\pm$0.002 & 0.025$\pm$0.002 & 157.3$\pm$1.3 \\
2013-03-02 & C1.9 & 14:59-15:11-15:26 & [-1200, -900, 0, 300] & (-956.8, 167.1) & (-1018.6, 85.1) & 74.58 & 0.063$\pm$0.004 & 0.063$\pm$0.003 & 0.065$\pm$0.005 & 158.9$\pm$0.8 \\
2015-01-30 & C2.0 & 17:43-18:06-18:20 & [300, 600, -170, 130] & (438.9, -94.3) & (492.3, 62.8) & 120.54 & 0.083$\pm$0.004 & 0.076$\pm$0.005 & 0.080$\pm$0.006 & 210.8$\pm$1.0 \\
2015-01-27 & C2.1 & 07:13-07:33-08:27 & [-1180, -880, -30, 270] & (-977.2, 112.9) & (-1103.4, 152.2) & 96.02 & 0.047$\pm$0.004 & 0.053$\pm$0.003 & 0.050$\pm$0.003 & 195.1$\pm$1.3 \\
2015-02-09 & C2.1 & 01:26-01:48-02:02 & [300, 600, -170, 130] & (427.8, -15.7) & (342.6, 22.5) & 67.87 & 0.049$\pm$0.005 & 0.047$\pm$0.005 & 0.052$\pm$0.004 & 177.9$\pm$1.5 \\
2016-08-31 & C2.2 & 20:10-20:19-20:39 & [-1170, -870, 80, 380] & (-897.7, 173.7) & (-935.2, 339.2) & 123.29 & 0.053$\pm$0.003 & 0.055$\pm$0.004 & 0.053$\pm$0.005 & 156.9$\pm$0.9 \\
2015-03-02 & C2.3 & 02:34-03:01-03:07 & [840, 1140, 200, 500] & (902.6, 356.7) & (1055.3, 267.8) & 128.35 & 0.048$\pm$0.003 & 0.064$\pm$0.004 & 0.047$\pm$0.004 & 181.8$\pm$1.0 \\
2015-01-03 & C2.4 & 07:05-07:15-07:28 & [-550, -250, -120, 180] & (-320.3, -38.0) & (-452.9, 105.0) & 141.69 & 0.042$\pm$0.003 & 0.045$\pm$0.002 & 0.041$\pm$0.003 & 179.7$\pm$0.9 \\
2014-01-07 & C2.4 & 04:40-04:53-05:20 & [-300, 0, -100, 200] & (-82.5, -70.8) & (-98.8, 48.6) & 87.48 & 0.030$\pm$0.002 & $\dots$ & 0.030$\pm$0.003 & 159.3$\pm$1.5 \\
2013-02-24 & C2.6 & 13:17-14:35-14:57 & [900, 1200, -30, 270] & (972.9, 99.9) & (1039.2, 26.2) & 72.03 & $\dots$ & $\dots$ & 0.061$\pm$0.003 & 166.3$\pm$0.6 \\
2015-02-08 & C2.6 & 10:32-11:00-11:16 & [140, 440, -170, 130] & (280.4, -8.1) & (296.5, 62.0) & 52.28 & 0.060$\pm$0.006 & 0.067$\pm$0.004 & 0.065$\pm$0.004 & 157.6$\pm$1.0 \\
2016-09-22 & C2.6 & 19:05-19:16-19:45 & [880, 1180, 40, 340] & (950.5, 199.6) & (1030.1, 115.7) & 84.04 & 0.069$\pm$0.006 & 0.065$\pm$0.006 & 0.071$\pm$0.005 & 163.4$\pm$1.0 \\
2016-02-19 & C2.7 & 22:49-23:10-23:18 & [-1130, -830, -550, -250] & (-896.0, -398.0) & (-975.7, -328.1) & 77.07 & 0.056$\pm$0.003 & 0.060$\pm$0.004 & 0.065$\pm$0.003 & 176.6$\pm$0.7 \\
2015-02-10 & C2.7 & 08:29-08:44-08:53 & [-930, -630, 160, 460] & (-870.8, 208.7) & (-842.9, 324.9) & 86.75 & 0.068$\pm$0.003 & 0.043$\pm$0.003 & 0.044$\pm$0.004 & 229.1$\pm$1.3 \\
2015-01-20 & C2.9 & 22:57-23:18-23:36 & [-1150, -850, -400, -100] & (-967.4, -194.4) & (-1018.7, -167.5) & 42.14 & 0.040$\pm$0.004 & 0.047$\pm$0.003 & 0.048$\pm$0.003 & 183.8$\pm$1.0 \\
2016-01-21 & C2.9 & 01:17-01:45-01:53 & [-760, -460, -230, 70] & (-563.0, -150.4) & (-670.7, -186.2) & 82.46 & 0.058$\pm$0.003 & 0.048$\pm$0.005 & 0.058$\pm$0.003 & 217.9$\pm$1.3 \\
2011-03-10 & C2.9 & 03:50-03:58-04:17 & [0, 300, 140, 440] & (115.9, 255.0) & (214.0, 414.3) & 135.92 & 0.063$\pm$0.004 & 0.070$\pm$0.004 & 0.052$\pm$0.004 & 182.3$\pm$0.9 \\
2016-06-13 & C3.0 & 05:33-05:52-06:18 & [850, 1150, 70, 370] & (929.4, 241.0) & (973.9, 144.2) & 77.41 & 0.045$\pm$0.004 & 0.047$\pm$0.004 & 0.051$\pm$0.003 & 177.8$\pm$1.2 \\
2014-01-03 & C3.0 & 06:29-07:00-07:18 & [-1050, -750, -150, 150] & (-844.5, -102.2) & (-919.4, 25.5) & 107.57 & 0.030$\pm$0.003 & 0.038$\pm$0.005 & 0.033$\pm$0.004 & 183.4$\pm$2.0 \\
2015-02-08 & C3.1 & 20:23-21:04-21:16 & [240, 540, -170, 130] & (362.9, -45.9) & (309.8, 17.6) & 60.13 & 0.090$\pm$0.004 & 0.046$\pm$0.004 & 0.090$\pm$0.008 & 186.0$\pm$1.1 \\
2016-02-27 & C3.3 & 05:44-05:55-06:01 & [-470, -170, -90, 210] & (-359.2, -30.7) & (-369.5, 45.6) & 55.85 & 0.049$\pm$0.005 & 0.033$\pm$0.005 & $\dots$ & 132.9$\pm$1.6 \\
2012-03-05 & C3.4 & 00:23-00:36-00:47 & [-920, -620, 240, 540] & (-773.9, 353.5) & (-861.9, 364.1) & 64.35 & $\dots$ & 0.053$\pm$0.005 & $\dots$ & 243.4$\pm$2.3 \\
2016-07-21 & C3.7 & 00:19-00:37-00:41 & [500, 800, -60, 240] & (622.9, -13.5) & (557.4, 79.8) & 82.78 & 0.056$\pm$0.004 & 0.054$\pm$0.004 & 0.054$\pm$0.003 & 182.3$\pm$1.0 \\
2012-03-10 & C3.8 & 00:19-01:19-01:45 & [-1100, -800, 160, 460] & (-864.3, 305.1) & (-985.5, 328.3) & 89.71 & 0.072$\pm$0.004 & 0.064$\pm$0.006 & $\dots$ & 252.0$\pm$1.9 \\
2016-04-15 & C4.1 & 14:04-14:30-14:43 & [220, 520, 40, 340] & (258.1, 275.5) & (278.6, 147.1) & 94.50 & $\dots$ & $\dots$ & 0.046$\pm$0.004 & 296.6$\pm$3.2 \\
2016-10-17 & C4.2 & 00:08-00:38-00:52 & [850, 1150, -440, -140] & (947.7, -249.7) & (1006.0, -391.3) & 111.20 & 0.061$\pm$0.004 & 0.059$\pm$0.002 & $\dots$ & 213.0$\pm$1.6 \\
2011-02-13 & C4.7 & 13:44-13:56-14:11 & [-250, 50, -370, -70] & (-126.4, -225.1) & (-91.3, -147.8) & 61.68 & 0.052$\pm$0.003 & 0.043$\pm$0.003 & 0.052$\pm$0.004 & 166.1$\pm$0.8 \\
\end{longtable}
\end{sidewaystable}

\begin{sidewaystable}
\begin{longtable}{l l l l l l l l l l l}
\toprule
Event & Flare class & Flare time (UT) & Observation region & Flare centre & Slice centre & Proj. dist. & $A_{\rm pre}$ & $A_{\rm during}$ & $A_{\rm post}$ & $P_{\rm mean}$ \\
& & & (arcsec) & (arcsec) & (arcsec) & (Mm) & (Mm) & (Mm) & (Mm) & (s) \\
\midrule
\endfirsthead
\bottomrule
\endlastfoot
2015-01-28 & C4.7 & 00:55-01:28-01:48 & [-1200, -900, -40, 260] & (-958.3, 131.8) & (-1056.3, 195.5) & 84.97 & 0.049$\pm$0.004 & 0.053$\pm$0.003 & 0.054$\pm$0.005 & 140.0$\pm$0.8 \\
2014-02-03 & C4.9 & 04:45-05:35-05:58 & [-160, 140, 220, 520] & (-67.2, 271.0) & (-52.0, 480.7) & 152.77 & $\dots$ & 0.044$\pm$0.005 & 0.036$\pm$0.003 & 167.9$\pm$1.7 \\
2011-04-16 & C5.2 & 00:34-00:57-01:08 & [250, 550, 200, 500] & (440.8, 303.1) & (397.6, 252.2) & 48.51 & 0.050$\pm$0.003 & 0.050$\pm$0.004 & 0.052$\pm$0.002 & 186.1$\pm$0.9 \\
2012-01-27 & C5.5 & 06:24-06:42-06:53 & [720, 1020, 480, 780] & (822.4, 511.0) & (899.7, 515.6) & 56.27 & 0.046$\pm$0.003 & 0.047$\pm$0.006 & $\dots$ & 191.3$\pm$1.9 \\
2015-01-06 & C6.0 & 05:16-05:31-05:58 & [200, 500, -100, 200] & (280.4, -75.4) & (267.5, 85.4) & 117.20 & $\dots$ & 0.042$\pm$0.003 & 0.040$\pm$0.004 & 144.2$\pm$1.1 \\
2014-01-27 & C6.0 & 08:07-08:23-08:40 & [-1200, -900, -50, 250] & (-981.1, 126.0) & (-1106.4, 161.0) & 94.47 & 0.031$\pm$0.002 & 0.038$\pm$0.004 & 0.043$\pm$0.003 & 166.9$\pm$1.1 \\
2014-01-05 & C6.6 & 15:11-15:18-15:31 & [-650, -350, -120, 180] & (-492.8, -28.3) & (-484.7, 109.7) & 100.47 & 0.041$\pm$0.003 & 0.041$\pm$0.004 & 0.039$\pm$0.003 & 234.7$\pm$2.1 \\
2015-01-29 & C8.2 & 05:15-05:23-05:50 & [-20, 280, -170, 130] & (55.0, -126.2) & (71.3, 96.2) & 161.95 & 0.040$\pm$0.003 & 0.045$\pm$0.004 & 0.043$\pm$0.006 & 164.5$\pm$1.3 \\
2014-01-28 & C8.5 & 16:23-16:45-16:57 & [-1180, -880, -400, -100] & (-977.6, -212.4) & (-990.7, -363.8) & 110.42 & 0.070$\pm$0.004 & 0.091$\pm$0.009 & 0.068$\pm$0.004 & 210.2$\pm$1.0 \\
2015-01-28 & C9.8 & 05:20-05:30-05:38 & [-250, 50, -140, 160] & (-147.1, -82.1) & (-132.1, 19.6) & 74.74 & 0.042$\pm$0.003 & 0.046$\pm$0.005 & 0.042$\pm$0.003 & 164.9$\pm$1.1 \\
2012-02-06 & M1.0 & 19:31-20:00-20:17 & [750, 1050, 170, 470] & (898.1, 246.9) & (924.0, 392.4) & 107.37 & 0.071$\pm$0.003 & 0.068$\pm$0.004 & $\dots$ & 199.6$\pm$0.8 \\
2011-03-25 & M1.0 & 23:08-23:22-23:30 & [-700, -400, -170, 130] & (-413.1, -104.7) & (-583.8, 7.7) & 148.47 & $\dots$ & 0.050$\pm$0.005 & 0.050$\pm$0.003 & 172.3$\pm$1.6 \\
2012-10-10 & M1.0 & 04:51-05:04-05:20 & [-1100, -800, -590, -290] & (-865.2, -436.0) & (-951.4, -435.0) & 62.64 & $\dots$ & 0.067$\pm$0.014 & 0.064$\pm$0.003 & 177.4$\pm$2.6 \\
2011-02-16 & M1.0 & 01:32-01:39-01:46 & [260, 560, -460, -160] & (354.7, -233.9) & (391.3, -321.1) & 68.69 & 0.052$\pm$0.005 & 0.045$\pm$0.003 & 0.043$\pm$0.005 & 175.7$\pm$1.8 \\
2012-08-17 & M1.0 & 17:08-17:20-17:27 & [-1130, -830, 160, 460] & (-897.2, 347.8) & (-979.2, 227.5) & 105.77 & 0.050$\pm$0.004 & 0.053$\pm$0.004 & 0.050$\pm$0.004 & 196.4$\pm$1.2 \\
2012-07-07 & M1.0 & 08:18-08:28-08:39 & [-1100, -800, -450, -150] & (-874.9, -282.2) & (-904.5, -362.2) & 62.00 & 0.039$\pm$0.002 & 0.051$\pm$0.002 & 0.049$\pm$0.003 & 188.4$\pm$0.9 \\
2011-02-16 & M1.1 & 07:35-07:44-07:55 & [330, 630, -450, -150] & (447.7, -243.4) & (417.8, -380.0) & 101.59 & 0.095$\pm$0.003 & 0.096$\pm$0.003 & 0.097$\pm$0.004 & 196.7$\pm$0.6 \\
2011-08-03 & M1.1 & 03:08-03:37-03:51 & [260, 560, 20, 320] & (410.3, 164.9) & (474.6, 172.3) & 46.97 & 0.034$\pm$0.005 & 0.038$\pm$0.003 & $\dots$ & 141.3$\pm$0.9 \\
2011-02-28 & M1.1 & 12:38-12:52-13:03 & [-830, -530, 360, 660] & (-583.1, 520.7) & (-690.6, 529.8) & 78.35 & 0.053$\pm$0.005 & $\dots$ & $\dots$ & 156.4$\pm$1.5 \\
2011-04-22 & M1.2 & 15:47-15:53-16:11 & [-750, -450, -310, -10] & (-527.5, -224.7) & (-614.0, -136.0) & 89.98 & 0.057$\pm$0.002 & 0.053$\pm$0.005 & 0.053$\pm$0.003 & 195.6$\pm$0.9 \\
2012-09-09 & M1.2 & 21:50-22:36-22:56 & [650, 950, -450, -150] & (759.8, -306.4) & (753.8, -406.5) & 72.81 & 0.053$\pm$0.005 & 0.055$\pm$0.003 & $\dots$ & 153.5$\pm$0.7 \\
2012-07-05 & M1.2 & 13:05-13:18-13:32 & [370, 670, -550, -250] & (606.3, -305.1) & (492.6, -459.5) & 139.27 & 0.072$\pm$0.003 & 0.069$\pm$0.004 & 0.060$\pm$0.005 & 197.1$\pm$1.0 \\
2011-03-07 & M1.2 & 05:00-05:13-05:19 & [600, 900, 380, 680] & (640.4, 492.2) & (772.3, 541.9) & 102.39 & 0.035$\pm$0.002 & 0.051$\pm$0.004 & 0.056$\pm$0.003 & 159.7$\pm$0.8 \\
2013-01-11 & M1.2 & 08:43-09:11-09:17 & [-770, -470, -70, 230] & (-588.8, 151.9) & (-514.5, 105.4) & 63.69 & 0.054$\pm$0.004 & 0.053$\pm$0.004 & 0.054$\pm$0.003 & 154.3$\pm$0.7 \\
2015-01-04 & M1.3 & 15:18-15:36-15:53 & [-260, 40, -300, 0] & (-67.2, -23.7) & (-121.6, -189.5) & 126.72 & 0.026$\pm$0.004 & 0.033$\pm$0.003 & 0.033$\pm$0.003 & 142.7$\pm$1.3 \\
2011-03-12 & M1.3 & 04:33-04:43-04:48 & [430, 730, 110, 410] & (538.6, 214.3) & (469.4, 288.2) & 73.51 & 0.050$\pm$0.003 & 0.054$\pm$0.004 & 0.049$\pm$0.004 & 152.7$\pm$1.0 \\
2012-10-21 & M1.3 & 19:46-20:03-20:20 & [-1100, -800, -490, -190] & (-923.5, -216.7) & (-954.7, -382.2) & 122.34 & 0.059$\pm$0.003 & 0.054$\pm$0.005 & 0.056$\pm$0.004 & 190.5$\pm$1.1 \\
2012-09-08 & M1.4 & 17:35-17:59-18:20 & [450, 750, -500, -200] & (588.7, -324.0) & (584.3, -406.5) & 60.03 & 0.036$\pm$0.003 & 0.050$\pm$0.003 & 0.036$\pm$0.002 & 172.0$\pm$1.0 \\
2011-03-07 & M1.4 & 07:59-08:07-08:15 & [600, 900, 380, 680] & (634.7, 512.8) & (790.5, 518.4) & 113.31 & 0.041$\pm$0.002 & 0.045$\pm$0.003 & 0.046$\pm$0.004 & 153.2$\pm$0.9 \\
2012-05-05 & M1.4 & 13:19-13:23-13:29 & [-1150, -850, 70, 370] & (-936.2, 197.1) & (-991.6, 149.6) & 52.94 & 0.067$\pm$0.005 & 0.067$\pm$0.006 & 0.069$\pm$0.003 & 249.1$\pm$1.8 \\
2013-08-12 & M1.5 & 10:21-10:41-10:47 & [-450, -150, -640, -340] & (-293.3, -440.8) & (-211.5, -488.3) & 68.67 & 0.038$\pm$0.002 & 0.038$\pm$0.002 & 0.038$\pm$0.004 & 153.9$\pm$0.9 \\
2012-07-05 & M1.6 & 20:09-20:14-20:28 & [360, 660, -560, -260] & (549.3, -335.4) & (544.4, -444.8) & 79.58 & 0.048$\pm$0.005 & 0.060$\pm$0.008 & $\dots$ & 150.2$\pm$1.3 \\
2011-09-05 & M1.6 & 04:08-04:28-04:32 & [770, 1070, 170, 470] & (912.2, 304.7) & (954.4, 384.7) & 65.75 & 0.052$\pm$0.005 & 0.052$\pm$0.006 & 0.053$\pm$0.003 & 144.6$\pm$1.0 \\
2014-03-20 & M1.7 & 03:42-03:56-04:08 & [-1160, -860, -400, -100] & (-916.8, -171.1) & (-1003.5, -261.6) & 91.00 & 0.039$\pm$0.003 & 0.056$\pm$0.005 & 0.042$\pm$0.002 & 260.6$\pm$1.5 \\
2011-09-21 & M1.8 & 12:04-12:23-12:45 & [-1130, -830, 200, 500] & (-932.0, 253.2) & (-991.3, 282.3) & 47.99 & 0.058$\pm$0.006 & 0.057$\pm$0.008 & 0.054$\pm$0.006 & 163.9$\pm$2.1 \\
\end{longtable}
\end{sidewaystable}

\begin{sidewaystable}
\begin{longtable}{l l l l l l l l l l l}
\toprule
Event & Flare class & Flare time (UT) & Observation region & Flare centre & Slice centre & Proj. dist. & $A_{\rm pre}$ & $A_{\rm during}$ & $A_{\rm post}$ & $P_{\rm mean}$ \\
& & & (arcsec) & (arcsec) & (arcsec)& (Mm) & (Mm) & (Mm) & (Mm) & (s) \\
\midrule
\endfirsthead
\bottomrule
\endlastfoot
2012-06-06 & M2.1 & 19:54-20:06-20:13 & [-100, 200, -430, -130] & (78.2, -289.6) & (29.3, -240.4) & 50.30 & 0.049$\pm$0.005 & 0.049$\pm$0.003 & 0.050$\pm$0.004 & 173.7$\pm$1.1 \\
2012-07-04 & M2.3 & 12:07-12:24-12:32 & [80, 380, -520, -220] & (256.0, -365.7) & (200.2, -445.8) & 70.89 & 0.055$\pm$0.003 & 0.057$\pm$0.004 & 0.052$\pm$0.006 & 180.8$\pm$1.1 \\
2013-08-17 & M3.3 & 18:16-18:24-18:35 & [330, 630, -390, -90] & (484.0, -174.0) & (407.9, -288.6) & 99.94 & 0.074$\pm$0.004 & 0.078$\pm$0.002 & 0.074$\pm$0.004 & 203.3$\pm$0.8 \\
2013-04-12 & M3.3 & 19:52-20:38-20:46 & [500, 800, 300, 600] & (599.4, 388.3) & (575.4, 506.1) & 87.35 & $\dots$ & 0.032$\pm$0.003 & 0.037$\pm$0.004 & 154.9$\pm$1.3 \\
2012-11-21 & M3.5 & 15:10-15:30-15:38 & [-450, -150, -100, 200] & (-227.1, 144.1) & (-323.4, 68.1) & 89.18 & 0.054$\pm$0.003 & 0.091$\pm$0.004 & 0.054$\pm$0.003 & 155.3$\pm$0.7 \\
2011-02-24 & M3.5 & 07:23-07:35-07:42 & [-1160, -860, 120, 420] & (-934.1, 278.2) & (-1013.2, 298.7) & 59.33 & 0.067$\pm$0.007 & 0.047$\pm$0.005 & 0.062$\pm$0.004 & 162.3$\pm$0.8 \\
2012-10-22 & M5.0 & 18:38-18:51-19:01 & [-1100, -800, -530, -230] & (-842.5, -257.7) & (-973.7, -301.4) & 100.45 & 0.035$\pm$0.002 & 0.061$\pm$0.003 & 0.033$\pm$0.002 & 158.5$\pm$0.7 \\
2011-09-06 & M5.3 & 01:35-01:50-02:05 & [-120, 180, -40, 260] & (111.8, 112.8) & (-15.4, 47.3) & 103.94 & 0.052$\pm$0.003 & 0.049$\pm$0.006 & 0.059$\pm$0.005 & 181.3$\pm$1.3 \\
2013-06-07 & M5.9 & 22:11-22:49-23:04 & [800, 1100, -530, -230] & (810.5, -494.4) & (964.4, -330.8) & 163.14 & $\dots$ & 0.053$\pm$0.004 & 0.051$\pm$0.003 & 212.2$\pm$1.3 \\
2011-08-03 & M6.0 & 13:17-13:48-14:10 & [320, 620, 10, 310] & (446.8, 207.5) & (536.2, 241.1) & 69.40 & 0.077$\pm$0.005 & 0.092$\pm$0.005 & 0.091$\pm$0.007 & 185.0$\pm$1.0 \\
2012-07-28 & M6.1 & 20:44-20:56-21:04 & [-910, -610, -500, -200] & (-707.3, -420.4) & (-835.6, -317.8) & 119.25 & 0.047$\pm$0.006 & 0.049$\pm$0.003 & 0.049$\pm$0.004 & 170.5$\pm$1.2 \\
2011-09-08 & M6.7 & 15:32-15:46-15:52 & [450, 750, -20, 280] & (608.4, 142.2) & (555.3, 70.5) & 64.79 & 0.051$\pm$0.003 & 0.054$\pm$0.005 & 0.050$\pm$0.004 & 150.5$\pm$0.9 \\
2013-11-08 & X1.1 & 04:20-04:26-04:29 & [-400, -100, -360, -60] & (-240.1, -282.6) & (-377.7, -144.7) & 141.49 & 0.077$\pm$0.002 & 0.072$\pm$0.004 & 0.070$\pm$0.004 & 229.4$\pm$1.0 \\
2014-01-07 & X1.2 & 18:04-18:32-18:58 & [-50, 250, -180, 120] & (118.3, -133.7) & (109.4, -60.0) & 53.94 & 0.035$\pm$0.004 & 0.028$\pm$0.003 & 0.028$\pm$0.004 & 156.8$\pm$1.6 \\
2011-03-09 & X1.5 & 23:13-23:23-23:29 & [10, 310, 140, 440] & (170.7, 275.4) & (159.3, 166.5) & 79.57 & 0.050$\pm$0.005 & 0.040$\pm$0.006 & 0.050$\pm$0.005 & 160.2$\pm$1.1 \\
2014-10-22 & X1.6 & 14:02-14:28-14:50 & [-470, -170, -550, -250] & (-255.7, -310.5) & (-382.0, -459.8) & 141.98 & 0.047$\pm$0.003 & 0.034$\pm$0.003 & 0.048$\pm$0.002 & 178.0$\pm$0.9 \\
2014-11-07 & X1.6 & 16:53-17:26-17:34 & [-750, -450, 110, 410] & (-597.7, 230.7) & (-579.7, 342.3) & 82.17 & 0.047$\pm$0.007 & 0.045$\pm$0.007 & 0.052$\pm$0.005 & 192.5$\pm$1.7 \\
2014-10-24 & X3.1 & 21:07-21:41-22:13 & [30, 330, -560, -260] & (281.1, -329.7) & (284.3, -466.5) & 99.45 & 0.025$\pm$0.002 & 0.026$\pm$0.003 & $\dots$ & 146.9$\pm$1.3 \\
2017-09-10 & X8.2 & 15:35-16:06-16:31 & [820, 1120, -450, -150] & (946.3, -205.3) & (928.8, -362.7) & 115.02 & 0.068$\pm$0.003 & 0.111$\pm$0.005 & 0.064$\pm$0.004 & 169.6$\pm$0.6 \\
2017-09-06 & X9.3 & 11:53-12:02-12:10 & [430, 730, -520, -220] & (536.3, -277.3) & (519.5, -412.0) & 98.56 & 0.054$\pm$0.006 & $\dots$ & 0.054$\pm$0.003 & 192.9$\pm$1.6 \\
\end{longtable}
\end{sidewaystable}
\FloatBarrier

\bibliography{declkink}
\bibliographystyle{aasjournalv7}
\end{document}